\documentclass[useAMS,usenatbib,usedcolumn]{mn2e}

\def\HI{\ifmmode{\rm HI}\else{H\/{\sc i}}\fi}

\def\HII{\hbox{\rm H\/{\sc ii}}\ }

\def\Ha{\hbox{\rm H$\alpha$}\ }

\def\lsun{\ifmmode{{\mathrm L}_{\odot}}\else{L$_{\odot}$}\fi} 

\def\deg{\hbox{$^\circ$}}
\def\arcmin{\hbox{$^\prime$}}
\def\arcsec{\hbox{$^{\prime\prime}$}}

\def\msun{\ifmmode{{\mathrm M}_{\odot}}\else{M$_{\odot}$}\fi} 
\def\msunpc2{\ifmmode{{\mathrm M}_{\odot} \, {\mathrm{pc}}^{-2}}\else{M$_{\odot} \, {\mathrm {pc}}^{-2}$}\fi}

\def\la{\mathrel{\mathchoice {\vcenter{\offinterlineskip\halign{\hfil
$\displaystyle##$\hfil\cr<\cr\sim\cr}}}
{\vcenter{\offinterlineskip\halign{\hfil$\textstyle##$\hfil\cr
<\cr\sim\cr}}}
{\vcenter{\offinterlineskip\halign{\hfil$\scriptstyle##$\hfil\cr
<\cr\sim\cr}}}
{\vcenter{\offinterlineskip\halign{\hfil$\scriptscriptstyle##$\hfil\cr
<\cr\sim\cr}}}}}
\def\ga{\mathrel{\mathchoice {\vcenter{\offinterlineskip\halign{\hfil
$\displaystyle##$\hfil\cr>\cr\sim\cr}}}
{\vcenter{\offinterlineskip\halign{\hfil$\textstyle##$\hfil\cr
>\cr\sim\cr}}}
{\vcenter{\offinterlineskip\halign{\hfil$\scriptstyle##$\hfil\cr
>\cr\sim\cr}}}
{\vcenter{\offinterlineskip\halign{\hfil$\scriptscriptstyle##$\hfil\cr
>\cr\sim\cr}}}}}

\usepackage[figuresright]{rotating}
\usepackage{lscape}
\usepackage{graphics}
\usepackage{epsfig}
\usepackage{multirow}
\usepackage{bigdelim}
\usepackage{bigstrut}

\defcitealias{Noordermeer05}{Paper~I}
\defcitealias{Noordermeer06a}{Paper~II}

\title[Rotation curves of early-type disk galaxies]{The mass distribution in
  early-type disk galaxies: declining rotation curves and correlations with
  optical properties}
\author[E.~Noordermeer et al.]
  {E.~Noordermeer,$^{1,2}$\thanks{email:edo.noordermeer@nottingham.ac.uk} 
  J.~M.~van der Hulst,$^1$ R.~Sancisi,$^{1,3}$ R.~S.~Swaters$^4$ and 
  \newauthor 
  T.~S.~van Albada$^1$ \\ 
  $^1$Kapteyn Astronomical Institute, University of Groningen, PO Box 800,
  9700 AV Groningen, The Netherlands \\
  $^2$University of Nottingham, School of Physics and Astronomy, University
  Park, NG7 2RD Nottingham, UK \\
  $^3$INAF-Osservatorio Astronomico di Bologna, Via Ranzani 1, 40127 Bologna,
  Italy \\
  $^4$Department of Astronomy, University of Maryland, College Park, MD
  20742-2421, USA}

\begin{document}

\date{accepted for publication in MNRAS, 05-01-2007}

\maketitle

\begin{abstract}
  We present rotation curves for 19 early-type disk galaxies (S0 -- Sab). The 
  galaxies span a B-band absolute magnitude range from $-17.5$ to $-22$, but
  the majority have a high luminosity with ${\mathrm {M_B}} < -20$. Rotation
  velocities are measured from a combination of \HI\ velocity fields and
  long-slit optical emission line spectra along the major axis; the resulting
  rotation curves probe the gravitational potential on scales ranging from
  100~pc to 100~kpc. \\
  We find that the rotation curves generally rise rapidly in the central
  regions and often reach rotation velocities of 200~--~300~km/s within a few 
  hundred parsecs of the centre. The detailed shape of the central rotation
  curves shows a clear dependence on the concentration of the stellar light
  distribution and the bulge-to-disk luminosity ratio: galaxies with highly
  concentrated stellar light distributions reach the maximum in their rotation
  curves at relatively smaller radii than galaxies with small bulges and a
  relatively diffuse light distribution. We interpret this as a strong
  indication that the dynamics in the central regions are dominated by the
  stellar mass. \\
  At intermediate radii, many rotation curves decline, with the asymptotic
  rotation velocity typically 10~--~20\% lower than the maximum. The strength
  of the decline is correlated with the total luminosity of the galaxies, more
  luminous galaxies having on average more strongly declining rotation
  curves. At large radii, however, all declining rotation curves flatten out,
  indicating that substantial amounts of dark matter must be present in these
  galaxies too. \\    
  A comparison of our rotation curves with the Universal Rotation Curve from
  \citet{Persic96} reveals large discrepancies between the observed and
  predicted rotation curves; we argue that rotation curves form a
  multi-parameter family which is too complex to describe with a simple
  formula depending on total luminosity only. \\ 
  In a number of galaxies from our sample, there is evidence for the
  presence of rapidly rotating gas in the inner few hundred parsecs
  from the centers. The inferred central masses and mass densities are
  too high to be explained by the observed stellar components and
  suggest the presence of supermassive black holes in these galaxies. 
\end{abstract}


\section{Introduction}
\label{sec:introduction}
Rotation curves are the prime tool for studying the mass distribution in disk
galaxies. 
In normal, unperturbed galaxies, gas moves on circular orbits around the
centre, so measurements of the circular velocity can be used to yield the
enclosed mass at different radii.   
The study of the shapes of rotation curves therefore gives important insight
into the overall distribution of mass in disk galaxies.  
\HI\ rotation curves in particular are useful, because they probe the mass
distribution to much larger radii than can be achieved with optical data and
reach to the regions where dark matter dominates the gravitational potential.  
In fact, it was the discovery, first made in the 1970's
\citep{Rogstad72,Roberts75,Bosma78,Bosma81b}, that \HI\ rotation curves stay
flat till the last measured points, well outside the optical disk, which gave
the final, irrefutable evidence of the presence of large amounts of unseen
matter in galaxies \citep{Bosma81b, Albada85, Albada86, Begeman87}.

A long standing question concerns the relation between the shape of rotation
curves and other properties of individual galaxies.  
It has been known for a long time that the shape of a rotation curve is
strongly coupled to the optical luminosity of a galaxy: slowly rising and low
amplitude for low-luminosity galaxies, high central gradient and high rotation
velocities for high-luminosity systems \citep[e.g.][]{Rubin85, Burstein85}. 
However, the question of whether or not other optical properties influence
rotation curves as well has resulted in inconsistent answers. 
\citet{Rubin85} and \citet{Burstein85} found no dependence on morphological
type or on the shape of the light distribution (notably the bulge-to-disk
ratio) and presented synthetic rotation curves depending solely on a galaxy's
luminosity.  
This idea was later elaborated by \citet{Persic91} and \citet{Persic96}, who
presented a `universal rotation curve', an analytic formula describing the
shape of a rotation curve which only depends on total luminosity.

In contrast, several other studies suggested that rotation curve shape is not
correlated with luminosity only, but that other parameters need to be taken
into account as well.  
\citet{Corradi90} found that the shape of a rotation curve correlates with a
galaxy's morphological type: early-type spirals with large bulges have
rotation curves which rise more rapidly than galaxies of similar luminosity
but with a less concentrated light distribution.   
\citet{Casertano91} showed that the outer shape of rotation curves is
correlated with both the total luminosity and the shape of the light
distribution, exemplified by two luminous galaxies with highly concentrated
light distributions which have declining rotation curves.
\citet{Roscoe99} showed that the universal rotation curve formalism of
\citet{Persic96} can be improved by including surface brightness as parameter
influencing rotation curve shape. 
The dependence of rotation curve shape on the optical characteristics
was also confirmed in studies by e.g.\ \citet{Broeils92},
\citet{Swatersthesis}, \citet{Verheijen01a}, \citet{Matthews02} and
\citet{Sancisi04}.   

The systematics behind rotation curve shapes hold important clues on the
structure and evolution of (disk) galaxies. 
\citet{Rubin85} and \citet{Burstein85} interpreted the lack of dependence on
the light distribution as evidence that luminous matter plays a minor r\^ole
dynamically, and that large amounts of dark matter must be present everywhere
in disk galaxies. 
But if rotation curves are, instead, a multi-parameter family depending also
on properties such as morphological type, surface brightness, etc., then the
conclusion must be that at least in some galaxies, the stars contribute
significantly to the potential. 
The rotation curve vs.\ optical properties relations also provide a powerful
benchmark for simulations of galaxy formation: any viable theory of galaxy
formation must be able to reproduce realistic rotation curves which match the
other characteristics of the simulated galaxy. 

In order to obtain a better understanding of these issues, a systematic study
of \HI\ rotation curves in spiral galaxies, covering a large range of
luminosities, morphological types and surface brightnesses, is a crucial step.
Although much work has been done in this field in recent years  
\citep[e.g.][]{DeBlok96, Swatersthesis, Cote00, Verheijen01a,
Gentile04}, most studies have focused on late-type and low-luminosity 
galaxies. 
Early-type disk galaxies, which generally contain less gas \citep{Roberts94,
Noordermeer05}, have received considerably less attention. 
One of the few studies so far aimed at a systematic investigation of 
\HI\ rotation curves over the full range of morphological types was
that by \citet{Broeils92}. 
However, in his sample of 23 galaxies, only one was of morphological
type earlier than Sb and only four had $V_{\mathrm {max}} > 250 \,
{\mathrm {km \, s^{-1}}}$.   
The only large-scale \HI\ survey directed specifically at S0 and Sa
galaxies was carried out by \citet{VanDriel87}, but his study was
severely hampered by the low signal-to-noise ratio of his data and
his rotation curves were of rather poor quality compared to modern
standards.
In the optical, little work has been done on early-type spiral galaxies
either, since the early studies by \citet{Rubin85} and \citet{Kent88}.  
S0 and Sa galaxies were thus also under-represented in the study by 
\citet{Persic96}; their Universal Rotation Curve is based on over 
1000 rotation curves of which only 2 are of type Sab or earlier.  

This paper is part of a larger study designed to fill this lack and to
systematically investigate the relation between dark and luminous matter in
early-type disk galaxies. 
These systems, lying at the high mass, high surface brightness end of the disk
galaxy population, are ideal test cases to investigate what determines the
shape of rotation curves.  
If the stars contribute significantly to the gravitational potentials of
galaxies, it is in these galaxies that their influence will be most easily
detected. 
In an earlier paper \citep[][ hereafter paper~I]{Noordermeer05}, we have
presented \HI\ observations for a sample of early-type (S0 -- Sab) disk
galaxies, and in an accompanying paper to the present one \citep[][
paper~II]{Noordermeer06a} we present optical photometry and bulge-disk
decompositions. 
Here, we use the data for a subset of 19 galaxies from
\citetalias{Noordermeer05} to derive their rotation curves and to study the
dependence of their rotation curve shapes on the optical properties. 
In two future publications, we will use the results to study the location of
massive, early-type disk galaxies on the Tully-Fisher relation and to create
detailed mass-models. 

The \HI\ data from \citetalias{Noordermeer05} can be used to measure the
rotation velocities of the gas out to large radii.  
In the central regions, however, the rotation curves can often not be measured
from the 21cm observations due to the presence of holes in the \HI\ disks
\citepalias[see][]{Noordermeer05}.    
Furthermore, the spatial resolution of our \HI\ observations is usually
insufficient to obtain detailed information on the shape of the rotation
curves in the inner regions, where our velocity fields suffer from beam
smearing.  
To overcome these difficulties, we use long-slit optical spectroscopy to
measure the central rotation curves.  
In most galaxies, optical emission lines can be detected in the very
inner regions, out to radii where reliable rotation velocities can be
determined from the \HI\ velocity fields. 
Moreover, due to the higher spatial resolution of the optical
observations, the effects of beam smearing are strongly reduced. 
\begin{table*}
 \begin{minipage}{12.25cm}
  \centering
   \caption[Rotation curve sample galaxies: basic data]
   {Sample galaxies: basic data. (1)~sample number; (2)~UGC number;
    (3)~alternative name; (4)~morphological type; (5)~distance; (6)~and
    (7)~absolute B- and R-band magnitudes (corrected for Galactic foreground
    extinction); (8)~R-band central disk surface brightness (corrected for
    Galactic foreground extinction and inclination effects) and (9)~R-band
    disk scale length. Column (4) was taken from NED, (5) from
    \citetalias{Noordermeer05} and (6) -- (9) from
    \citetalias{Noordermeer06a}.   
    \label{table:sample}}  
  
   \begin{tabular}{r@{\hspace{0.6cm}}r@{\hspace{0.6cm}}llrrrrr}
    \hline 
    \multicolumn{1}{c}{sample} & \multicolumn{1}{c@{\hspace{0.6cm}}}{UGC} & 
    \multicolumn{1}{c}{\hspace{-0.2cm}alternative} &
    \multicolumn{1}{c}{Type} & \multicolumn{1}{c}{D} &
    \multicolumn{1}{c}{M$_{\mathrm B}$} & \multicolumn{1}{c}{M$_{\mathrm R}$}
    & \multicolumn{1}{c}{$\mu_{0,R}^c$} & \multicolumn{1}{c}{$h_R$} \\   
    
    \multicolumn{1}{c}{number} & & \multicolumn{1}{c}{\hspace{-0.2cm}name} &
     & \multicolumn{1}{c}{Mpc} & \multicolumn{1}{c}{mag} &
    \multicolumn{1}{c}{mag} &
    \multicolumn{1}{c}{$\frac{\mathrm{mag}}{\mathrm{arcsec}^2}$} & 
    \multicolumn{1}{c}{kpc} \\  
    
    \multicolumn{1}{c}{(1)} & \multicolumn{1}{c@{\hspace{0.6cm}}}{(2)} &
    \multicolumn{1}{c}{\hspace{-0.2cm}(3)} & \multicolumn{1}{c}{(4)} & 
    \multicolumn{1}{c}{(5)} & \multicolumn{1}{c}{(6)} &
    \multicolumn{1}{c}{(7)} & \multicolumn{1}{c}{(8)} &
    \multicolumn{1}{c}{(9)} \\  
    \hline 
    
     1 & 624   & NGC 338  & Sab          & 65.1 & -20.83                           & -22.25 & 21.92 & 5.8 \\
     2 & 2487  & NGC 1167 & SA0-         & 67.4 & -21.88                           & -23.24 & 20.12 & 8.0 \\ 
     3 & 2916  & --       & Sab          & 63.5 & -21.05                           & -22.01 & 20.99 & 5.0 \\
     4 & 2953  & IC 356   & SA(s)ab pec  & 15.1 & -21.22                           & -22.54 & 19.25 & 4.1 \\
     5 & 3205  & --       & Sab          & 48.7 & -20.89                           & -21.88 & 19.59 & 3.5 \\
     6 & 3546  & NGC 2273 & SB(r)a       & 27.3 & -20.02                           & -21.35 & 19.49 & 2.8 \\ 
     7 & 3580  & --       & SA(s)a pec:  & 19.2 & -18.31                           & -19.42 & 21.58 & 2.4 \\
     8 & 3993  & --       & S0?          & 61.9 & -20.19                           & -21.35 & 22.37 & 5.5 \\
     9 & 4458  & NGC 2599 & SAa          & 64.2 & -21.38                           & -22.61 & 21.26 & 8.6 \\ 
    10 & 4605  & NGC 2654 & SBab: sp     & 20.9 & -20.09$^\dagger$\hspace{-0.14cm} & --$^\dagger$ & --$^\dagger$ & --$^\dagger$  \\
    11 & 5253  & NGC 2985 & (R')SA(rs)ab & 21.1 & -20.86                           & -21.90 & 21.32 & 5.3 \\
    12 & 6786  & NGC 3900 & SA(r)0+      & 25.9 & -19.94$^\dagger$\hspace{-0.14cm} & -21.13 & 19.30 & 1.5 \\
    13 & 6787  & NGC 3898 & SA(s)ab      & 18.9 & -20.00                           & -21.28 & 20.49 & 3.3 \\
    14 & 8699  & NGC 5289 & (R)SABab:    & 36.7 & -19.48                           & -20.74 & 22.24 & 3.7 \\ 
    15 & 9133  & NGC 5533 & SA(rs)ab     & 54.3 & -21.22                           & -22.62 & 21.27 & 9.1 \\
    16 & 11670 & NGC 7013 & SA(r)0/a     & 12.7 & -19.20                           & -20.55 & 19.58 & 1.8 \\
    17 & 11852 & --       & SBa?         & 80.0 & -20.44                           & -21.53 & 20.74 & 4.5 \\
    18 & 11914 & NGC 7217 & (R)SA(r)ab   & 14.9 & -20.27                           & -21.35 & 19.91 & 2.7 \\
    19 & 12043 & NGC 7286 & S0/a         & 15.4 & -17.53                           & -18.26 & 19.90 & 0.8 \\ 
    \hline 
    \multicolumn{9}{l}{$^\dagger$ No data available in \citetalias{Noordermeer06a}; M$_{\mathrm B}$ taken from LEDA.}      
   \end{tabular}
 \end{minipage}
\end{table*}  

The remainder of this paper is structured as follows. 
In section~\ref{sec:sample}, the criteria which were used to 
select suitable galaxies from the parent sample of
\citetalias{Noordermeer05} are described.  
Section~\ref{sec:data} describes the techniques that were used to derive the
rotation curves from the \HI\ velocity fields and from the optical spectra. 
In section~\ref{sec:paramcomp}, the fitted orientation parameters and systemic
velocities of our galaxies, as derived from different sources, are compared.   
In section~\ref{sec:warps}, we briefly discuss the occurrence of warps in the
galaxies in our sample.  
In section~\ref{sec:rotcurshape}, several aspects of the shape of our rotation
curves are discussed, including an analysis of the correlations with optical
properties and the applicability of the concept of a `Universal Rotation
Curve' to our data.    
Finally, we briefly discuss our results and summarize the main conclusions in
section~\ref{sec:conclusions}.  
In the appendices, we present some additional material. 
A detailed description of the individual rotation curves is presented in
appendix~\ref{app:notes}. 
In appendix~\ref{app:rotcurs_centres}, we interpret the broad central velocity
profiles which are present in some of our optical spectra. 
Appendix~\ref{app:figures} gives the graphical representation of the rotation
curves and various other data for the galaxies in our sample.

\section{Sample selection}
\label{sec:sample}
The galaxies for the rotation curve study presented here were selected  
from the 68~galaxies with \HI\ observations presented in
\citetalias{Noordermeer05}, which were in turn selected from the WHISP 
survey \citep[Westerbork survey of \HI\ in spiral and irregular
galaxies;][]{Kamphuis96, VanderHulst01}.  
In order to be able to derive high quality \HI\ rotation curves, galaxies were
selected on the basis of the following criteria: {\em 1}) the velocity field
must be well resolved ($>$~5 -- 10 beams across) and defined over significant
parts of the gas disks (i.e.\ not confined to small `patches'); {\em 2}) the
gas must be moving in regular circular orbits around the centre of the galaxy. 
Strongly interacting galaxies, or galaxies with otherwise distorted kinematics
cannot be used.   
Strongly barred galaxies are excluded as well, because non-circular motions in
the bar potential complicate the analysis of the data; {\em 3}) the
inclination angle must be well constrained and preferably lie between 40\deg
and 80\deg.     

Few galaxies from \citetalias{Noordermeer05} satisfy all these conditions
and a strict application of these criteria (especially the second one) would
lead to a very small sample.   
We have therefore relaxed the latter two selection criteria and included a
number of galaxies with e.g.\ weak bars, mild kinematical distortions or a
more face- or edge-on orientation.    
The resulting sample consists of 19 galaxies; a few basic characteristics of
the members are given in table~\ref{table:sample}.   

The galaxies in our sample have morphological types ranging from S0-
to Sab and span two decades in optical luminosity ($-17.5 > {\mathrm
{M_B}} > -22$).  
The majority of galaxies in our sample have high optical luminosity,
with ${\mathrm {M_B}} < -20$. 
See \citetalias{Noordermeer05} for a more detailed description of the
properties of the galaxies in our sample. 

\section{Observations, data reduction and the derivation of the
     rotation curves}
\label{sec:data}
As mentioned in the introduction, the rotation curves in this paper
were derived from a combination of \HI\ synthesis observations and
long-slit optical spectra. 
Below, we discuss the analysis of both components separately.

\subsection{\HI\ rotation curves}
\label{subsec:HIrotcurs}
The \HI\ rotation curves were derived by fitting tilted ring models 
\citep{Begeman87,Begeman89} to the observed velocity fields from
\citetalias{Noordermeer05}, using the ROTCUR algorithm implemented in
GIPSY \citep[Groningen Image Processing System;][]{Vogelaar01}\footnote{For two
highly inclined galaxies, UGC~4605 and 8699, the standard tilted ring method
is not suitable and a modified analysis was applied (see individual notes in
appendix~\ref{app:notes}).}.    
In \citetalias{Noordermeer05}, we showed velocity fields at either full 
($\approx 15^{\prime\prime}$), 30\arcsec\ or 60\arcsec\
resolution. 
Here, we fit tilted ring models to the velocity fields at all available
resolutions.   
The higher-resolution velocity fields can be used for the inner regions,
whereas the velocity fields at lower resolution generally 
extend out to larger radii and can be used to obtain information about the
rotation curves in the outer parts.   
Tilted rings were fitted to the entire velocity fields, but points
near the major axis were given more weight than those near the minor 
axis by applying a $|\cos(\alpha)|$ weighing scheme, with $\alpha$ the  
azimuthal angle, measured from the major axis in the plane of the galaxy.  
\begin{table*}
 \begin{minipage}{16.5cm}
  \centering 
   \caption[Dynamical properties]
   {Dynamical properties: (1) UGC number; (2)~and (3)~position of
    the dynamical centre; (4) heliocentric systemic velocity;
    (5)~position angle (north through east) of major axis;
    (6)~inclination angle; (7)~maximum rotation velocity; (8)~rotation
    velocity at 2.2 R-band disk scale lengths; (9)~asymptotic rotation
    velocities at large radii; (10)~total enclosed mass within last
    measured point and (11)~rotation curve quality.  
    \label{table:data}} 

   \begin{tabular}{rr@{\hspace{0.2cm}}r@{\hspace{0.25cm}}d{2.1}@{\hspace{0.7cm}}r@{\hspace{0.25cm}}r@{\hspace{0.cm}}r@{\hspace{0.7cm}}rccr@{\hspace{0.31cm}}r@{\hspace{0.31cm}}r@{\hspace{0.68cm}}cc}
    \hline
    \multicolumn{1}{c}{UGC} & \multicolumn{6}{c}{dynamical centre} & 
    \multicolumn{1}{c}{$V_{\mathrm {sys}}$} & $PA$  &
    $i$ & \multicolumn{1}{c}{$V_{\mathrm {max}}$}
    & \multicolumn{1}{c}{$V_{\mathrm {2.2h}}$} &
    \multicolumn{1}{c@{\hspace{0.48cm}}}{$V_{\mathrm {asymp}}$} &  
    $M_{\mathrm {enc}}$ & quality \\    
    
     & \multicolumn{3}{c@{\hspace{0.7cm}}}{RA (2000)} &
    \multicolumn{3}{c@{\hspace{0.9cm}}}{Dec (2000)} & & & & & & & & \\ 
    
     & \multicolumn{1}{c}{\it h} & \multicolumn{1}{c}{\it m} &
    \multicolumn{1}{c@{\hspace{0.7cm}}}{\it s} &
    \multicolumn{1}{c}{$^{\circ}$} & \multicolumn{1}{c}{\arcmin} &
    \multicolumn{1}{c@{\hspace{0.7cm}}}{\arcsec} &
    \multicolumn{1}{c}{km/s} & \multicolumn{1}{c}{$^{\circ}$} & &
    \multicolumn{1}{c}{km/s} & \multicolumn{1}{c}{km/s} &
    \multicolumn{1}{c@{\hspace{0.48cm}}}{km/s} & \msun & \\ 
    
    \multicolumn{1}{c}{(1)} & \multicolumn{3}{c@{\hspace{0.7cm}}}{(2)} &  
    \multicolumn{3}{c@{\hspace{0.7cm}}}{(3)} & \multicolumn{1}{c}{(4)} & (5) &
    (6) & \multicolumn{1}{c}{(7)} & \multicolumn{1}{c}{(8)} &  
    \multicolumn{1}{c@{\hspace{0.48cm}}}{(9)} & (10) & (11) \\    
    \hline
    
      624 & 1  &  0 & 36.0 & 30 & 40 & 10 & 4789 & 288      & 64$^\dagger$\hspace{-0.13cm} & 300 & 300    & 270 & $5.2\cdot 10^{11}$ & III \\
     2487 & 3  &  1 & 42.7 & 35 & 12 & 21 & 4952 & 250--256 & 36           & 390 & 360    & 330 & $2.1\cdot 10^{12}$ & I   \\
     2916 & 4  & 2  & 33.5 & 71 & 42 & 19 & 4537 & 242      & 42--50       & 220 & 210    & 180 & $2.8\cdot 10^{11}$ & II  \\
     2953 & 4  &  7 & 46.8 & 69 & 48 & 46 & 892  & 98--104  & 50           & 315 & 315    & 260 & $1.1\cdot 10^{12}$ & I   \\ 
     3205 & 4  & 56 & 14.9 & 30 & 3  & 8  & 3586 & 230--224 & 67           & 240 & 230    & 210 & $4.3\cdot 10^{11}$ & I   \\
     3546 & 6  & 50 & 8.6  & 60 & 50 & 46 & 1837 & 56       & 55           & 260 & 185    & 190 & $2.4\cdot 10^{11}$ & II  \\
     3580 & 6  & 55 & 31.2 & 69 & 33 & 54 & 1203 & 6--356   & 63           & 127 & 100    & 125 & $9.0\cdot 10^{10}$ & II  \\
     3993 & 7  & 55 & 44   & 84 & 55 & 33 & 4364 & 220      & 20           & 300 & 290    & 250 & $8.6\cdot 10^{11}$ & II  \\
     4458 & 8  & 32 & 11.2 & 22 & 33 & 36 & 4756 & 288--295 & 25           & 490 & 280    & 240 & $7.8\cdot 10^{11}$ & II  \\
     4605 & 8  & 49 & 11.1 & 60 & 13 & 14 & 1347 & 240--247 & 84--74       & 225 & 220$^*$\hspace{-0.132cm} & 185 & $2.8\cdot 10^{11}$ & I   \\
     5253 & 9  & 50 & 22.2 & 72 & 16 & 44 & 1329 & 356--340 & 37           & 255 & 245    & 210 & $5.5\cdot 10^{11}$ & II  \\
     6786 & 11 & 49 & 9.2  & 27 & 1  & 15 & 1795 & 181--186 & 68--64       & 230 & --$^\$$\hspace{-0.132cm} & 215 & $3.1\cdot 10^{11}$ & I   \\
     6787 & 11 & 49 & 15.3 & 56 & 5  & 5  & 1172 & 107--118 & 69--66       & 270 & 250    & 250 & $5.0\cdot 10^{11}$ & I   \\
     8699 & 13 & 45 & 7.7  & 41 & 30 & 19 & 2516 & 280      & 73           & 205 & 190    & 180 & $1.9\cdot 10^{11}$ & I   \\
     9133 & 14 & 16 & 7.7  & 35 & 20 & 37 & 3858 & 24--45   & 53           & 300 & 265    & 225 & $1.3\cdot 10^{12}$ & I   \\
    11670 & 21 &  3 & 33.5 & 29 & 53 & 50 & 774  & 336--330 & 70--68       & 190 & 155    & 160 & $1.6\cdot 10^{11}$ & II  \\
    11852 & 21 & 55 & 59.6 & 27 & 53 & 55 & 5843 & 200--175 & 50--60       & 220 & 210    & 165 & $5.9\cdot 10^{11}$ & II  \\
    11914 & 22 & 7  & 52.3 & 31 & 21 & 36 & 951  & 265--268 & 31           & 305 & 300    & 300$^{\#}$\hspace{-0.23cm} & $1.9\cdot 10^{11}$ & II  \\
    12043 & 22 & 27 & 50.4 & 29 & 5  & 45 & 1007 & 97--92   & 67           & 93  &  82    & 90  & $3.2\cdot 10^{10}$ & I   \\
    \hline    
    \multicolumn{15}{l}{\small $^\dagger$ Kinematical inclination poorly constrained by observations. Value copied from optical 
      isophotal analysis.} \\
    \multicolumn{15}{l}{\small $^*$ No accurate photometry available due to
      edge-on orientation of optical disk; optical scale length is
      estimate only.} \\ 
    \multicolumn{15}{l}{\small $^\$$ Galaxy does not have regular exponential
      disk; no optical scale length available.} \\
    \multicolumn{15}{l}{\small $^{\#}$ Rotation curve extends out to 3.3 R
      disk scale lengths only and may converge to different velocity at larger
      radii.}  
   \end{tabular}
 \end{minipage}
\end{table*}  

In all cases, the rotation curves were determined in four steps. 
In the first step, all parameters (i.e.\ systemic velocity $V_{\mathrm 
{sys}}$, dynamical centre position $(x_c, y_c)$, position angle $PA$,
inclination angle $i$ and rotation velocity $V_{\mathrm {rot}}$) were
left free for each ring. 
In general, the fitted systemic velocities and dynamical centre
positions show little variation with radius, especially in the
inner regions, and the average values were adopted as the global
values for the galaxy. They are listed in table~\ref{table:data}.  

In the second step, the systemic velocity and dynamical centre were fixed for
each ring at the values derived in the first step.   
The values for the position angle derived from this step are shown with the
data points in the figures in appendix~\ref{app:figures}.
The position angle is usually well-defined, but it often shows variations with
radius as a result of warps in the gas disk.  
If a clear trend was visible, we fitted it by hand and used the fitted values
for the next steps; otherwise we used the average of all rings.
The adopted range of position angles, or the average value, for each galaxy is
given in table~\ref{table:data}, and plotted as bold line in the figures in
appendix~\ref{app:figures}. 

In the third step, only the inclination and rotation velocity were left as
free parameters for each ring. 
From this fit, the inclination angle was determined. 
This parameter is the most difficult one to constrain, because it is strongly
coupled to the rotation velocity, especially for inclinations lower than
$\approx 45$\deg\ \citep{Begeman87,Begeman89}.   
The fitted values are shown in the figures in appendix~\ref{app:figures}.   
It is clear that for the more face-on galaxies, the uncertainties in
the fitted inclinations are large. 
In practice, radial variations in inclination could only be detected
for galaxies that are sufficiently inclined; for galaxies with $i \la\ 
45^\circ$ only an average value could be determined. 
When necessary, the fitted inclination angles were also compared to the values
derived from the optical images \citepalias[see][]{Noordermeer06a} to make a
more reliable estimate.  
The range of inclination angles, or the average value, used for the next step
is given in table~\ref{table:data} and plotted as bold line in the figures in
appendix~\ref{app:figures}.   

The uncertainty in the inclination $\Delta i(r)$ was estimated by eye, based
on the spread of the individual data points around the fitted values and the
comparison between the tilted ring inclination angles and optical ellipticity. 
In general, we let the uncertainty $\Delta i$ increase with radius, in
order to account for the possibility of undetected or misfitted warps
in the outer gas disks of the galaxies. 
The adopted uncertainties in the inclination angle are shown with the
shaded regions in the bottom middle panels in the figures in
appendix~\ref{app:figures}.
\begin{table*}
 \begin{minipage}{13.8cm}
  \centering
   \caption[Observational parameters for optical spectroscopic observations] 
   {Observational parameters for optical spectroscopic observations:
    (1)~UGC number; (2)~telescope used: Isaac Newton Telescope (INT) or
    William Herschel Telescope (WHT) on La Palma or NOAO 2.1m telescope 
    on Kitt Peak (KP~2.1m); (3)~observing dates; (4)~total exposure
    time; (5)~and (6)~effective slit width and length; (7)~position
    angle (north through east) of the slit on the sky and (8)~--~(12)  
    line weights used in the stacking procedure. 
    \label{table:Haobservations}} 
  
   \begin{tabular}{rc@{\hspace{0.25cm}}rc@{\hspace{0.5cm}}c@{\hspace{0.1cm}}c@{\hspace{0.1cm}}r@{\hspace{0.6cm}}c@{\hspace{0.2cm}}c@{\hspace{0.2cm}}c@{\hspace{0.1cm}}c@{\hspace{0.1cm}}c@{\hspace{0.1cm}}} 
    \hline
    \multicolumn{1}{c}{UGC} & telescope & \multicolumn{1}{c}{dates} &   
    t$_{\mathrm {exp}}$ & \multicolumn{3}{c@{\hspace{0.75cm}}}{slit orientation} &
    \multicolumn{5}{c}{line weights $w_i$} \\  

     & & & & width & length & \multicolumn{1}{c@{\hspace{0.6cm}}}{PA} &  
    {\footnotesize [N{\sc ii}]$_{6548}$} & 
    {\footnotesize \Ha} &  
    {\footnotesize [N{\sc ii}]$_{6583}$} & 
    {\footnotesize [S{\sc ii}]$_{6716}$} & 
    {\footnotesize [S{\sc ii}]$_{6731}$} \\  

     & & & sec & \arcsec & \arcmin &
    \multicolumn{1}{c@{\hspace{0.6cm}}}{\deg} & & & & & \\   

    \multicolumn{1}{c}{(1)} & (2) & \multicolumn{1}{c}{(3)} & (4) & 
    (5) & (6) & (7) & (8) & (9) & (10) & (11) &
    (12) \\  

    \hline
    624   & INT     & 26/1/01 & 2400 & 1.5 & 3.3 & 106 & 0.5 & 2.0 & 1.0 & 0.5 & 0.5 \\
    2487  & INT     & 28/1/01 & 7200 & 1.0 & 3.3 &  70 & 0.5 & 1.0 & 1.0 & 0.5 & 0.5 \\ 
    2916  & INT     & 27/1/01 & 2400 & 1.0 & 3.0 &  76 & 0.0 & 1.0 & 1.0 & 0.0 & 0.0 \\
    2953  & WHT     & 2/1/00  & 7200 & 1.0 & 4.0 &  99 & 0.5 & 1.0 & 1.0 & 0.5 & 0.5 \\
    3205  & INT     & 26/1/01 & 6000 & 1.5 & 3.3 &  47 & 0.0 & 1.0 & 1.0 & 0.0 & 0.0 \\
    3546  & INT     & 26/1/01 & 3600 & 1.5 & 3.3 &  57 & 0.0 & 1.0 & 2.0 & 0.0 & 0.0 \\
    3580  & INT     & 28/1/01 & 3600 & 1.0 & 3.3 &   5 & 0.25& 1.0 & 0.5 & 0.5 & 0.5 \\
    3993  & INT     & 26/1/01 & 7200 & 1.5 & 3.0 &  44 & 0.25& 0.5 & 1.0 & 0.5 & 0.5 \\
    4458  & INT     & 27/1/01 & 3600 & 1.0 & 3.0 & 100 & 0.25& 1.0 & 1.0 & 0.25& 0.25\\
    4605  & INT     & 28/1/01 & 4800 & 1.0 & 3.3 &  63 & 0.0 & 2.0 & 1.0 & 0.5 & 0.5 \\
    5253  & INT     & 27/1/01 & 2400 & 1.0 & 3.2 &   0 & 0.1 & 1.0 & 1.0 & 0.3 & 0.3 \\
    6786  & INT     & 29/1/01 & 2400 & 1.0 & 3.3 &   2 & 0.0 & 1.0$^{\dagger}$\hspace{-0.12cm} & 1.0$^{\dagger}$\hspace{-0.12cm} & 0.0 & 0.0 \\ 
    6787  & INT     & 31/1/01 & 6000 & 1.0 & 3.3 & 105 & 0.1 & 1.0 & 0.5 & 0.25& 0.25\\
    8699  & INT     & 22/5/01 & 3600 & 1.0 & 3.6 & 100 & 0.25& 1.0 & 1.0 & 0.5 & 0.5 \\
    9133  & INT     & 22/5/01 & 7200 & 1.0 & 3.6 &  27 & 0.0 & 1.0 & 1.0 & 0.0 & 0.0 \\
    11670 & INT     & 22/5/01 & 4200 & 1.0 & 3.6 & 157 & 0.25& 1.0 & 1.0 & 0.5 & 0.5 \\
    11852 & INT     & 23/5/01 & 4800 & 1.0 & 3.6 &  15 & 0.0 & 1.0 & 1.0 & 0.0 & 0.0 \\
    11914 & INT     & 23/5/01 & 3150 & 1.0 & 3.6 &  89 & 0.25& 1.0 & 1.0 & 0.25& 0.25\\
    12043 & KP 2.1m & 8/12/01 & 2400 & 1.0 & 5.2 &  98 & 0.25& 2.0 & 0.5 & 0.7 & 0.0 \\    
    \hline
    \multicolumn{12}{l}{$^\dagger$ Lines could not be stacked because of
    stellar absorption feature in \Ha (see note in appendix~\ref{app:notes}).}
   \end{tabular}
 \end{minipage}
\end{table*}  

In the final step, we derived the rotation curves by doing a fit with
all parameters fixed except the rotation velocity.  
To prevent the inclusion of erroneous points in the final rotation curves,
outer tilted rings were not accepted if they only covered a small number of
pixels in the velocity field, or if the outer parts of the velocity fields
showed clear signs of non-circular or otherwise perturbed motions.  
The final \HI\ rotation curves are shown as square data points in the
bottom right panels in the figures in appendix~\ref{app:figures}.

\subsection{Rotation curves from optical spectra}
\label{subsec:Harotcurs}

\subsubsection{observations}
For most galaxies in the sample, long-slit spectra were taken with the
IDS spectrograph on the Isaac Newton Telescope (INT) on La Palma\footnote{The  
  Isaac Newton Telescope and William Herschel Telescope are operated on the
  island of La Palma by the Isaac Newton Group in the Spanish Observatorio del 
  Roque de los Muchachos of the Instituto de Astrofisica de Canarias.}.  
\addtocounter{footnote}{-1}
For UGC~2953, a spectrum was obtained from the red arm of the ISIS
spectrograph on the William Herschel Telescope, also on La Palma\footnotemark.
The spectrum of UGC~12043 was taken with the GoldCam spectrograph,
mounted on the NOAO 2.1m telescope on Kitt Peak, Arizona\footnote{The Kitt
  Peak 2.1m telescope is operated and IRAF is distributed by the National
  Optical Astronomy Observatories, which are operated by the Association of
  Universities for Research in Astronomy, Inc., under cooperative agreement
  with the National Science Foundation.}.    
\addtocounter{footnote}{-1}
A summary of the observations is given in table~\ref{table:Haobservations}.  

The slits of the spectrographs were aligned with the major axes of the
galaxies. 
In some cases, the position angle of the slit on the sky was slightly 
different from the kinematical position angle of the galaxy as
derived from the \HI\ velocity field. 
In these cases, the rotation curves were later corrected for the
effect of the misalignment. 
The bulges of the galaxies were usually bright enough to enable the
slit to be positioned accurately over the centres using the TV camera
in the focal plane. 

The spectral range of all observations was chosen such that each
spectrum contains the redshifted lines of \Ha ($\lambda_0 =
6562.80$~\AA), [N{\sc ii}] (6548.04 and 6583.46~\AA) and [S{\sc ii}]
(6716.44 and 6730.81~\AA). 
The spectral resolution of the spectra taken on the INT is 1.0 and
1.4~\AA\ (FWHM) for slit widths of 1.0 and 1.5\arcsec\ respectively, 
corresponding to a velocity resolution of approximately 45 and 65~km/s
respectively.   
The spectrum for UGC~2953 has a spectral resolution of 0.9~\AA\
($\sim$~40~km/s), whereas the resolution of the spectrum for UGC~12043
is slightly worse at 2.0~\AA\ ($\sim$~90~km/s). 

Total exposure times were broken up into single exposures of typically
20 minutes; the number of exposures for each galaxy was determined at
the telescope, based on the strength of the emission lines in the
first exposure. 

\subsubsection{data reduction}
\label{subsubsec:optredux}
Standard data reduction steps were performed within the IRAF
environment\footnotemark.   
Readout bias was subtracted using the overscan region of the chips;
any remaining structure was removed using special bias frames. 
The spectra were then flatfielded using Tungsten flatfields. 
Wavelength calibrations were performed using arc spectra from Copper, 
Neon and Argon lamps, taken before or after the galaxy spectra.   
The resulting wavelength solutions were used to map the spectra to a
logarithmic wavelength grid, such that pixel shifts correspond to linear
velocity shifts.  
The calibrations were also checked retrospectively by comparing the
measured wavelengths of a few strong night-sky lines with the values
given by \citet{Osterbrock96}; the systematic errors lie typically in the 
range 0.05 -- 0.10~\AA, corresponding to about 2.5 -- 5 km/s.  
\begin{figure*}
 \centerline{\psfig{figure=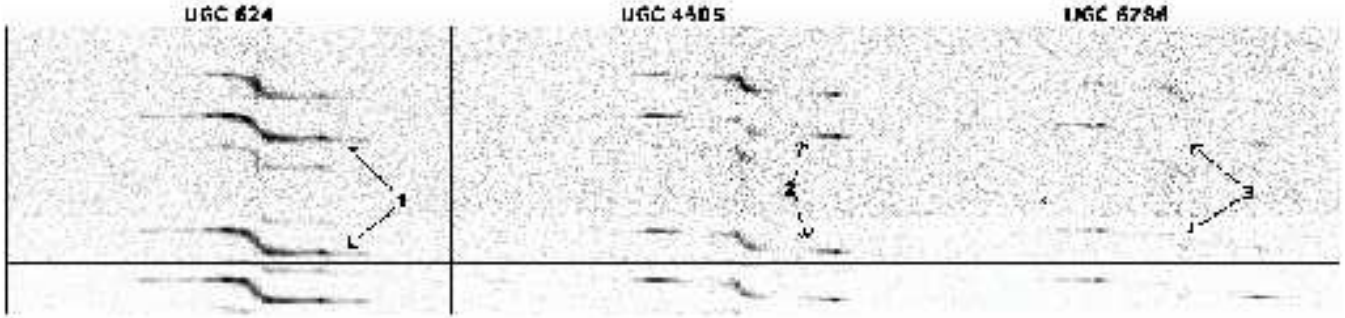,angle=270,width=17.8cm}}
  \caption{Example spectra to illustrate the spectral stacking procedure. The
    top panel at each column shows a section of the original spectrum,
    centered around the H$\alpha$- (middle) and the 6583 and 6548~\AA\ [N{\sc
    ii}]-lines (top and bottom respectively), after removal of the stellar
    continuum and sky lines. The middle panels show the stacked spectra, where
    the emission from all lines (including the [S{\sc ii}]-lines) is added,
    with weights $w_i$ as given in table~\ref{table:Haobservations}. The
    bottom panels show the same spectra after binning along the spatial
    direction to $\sim \! 1\arcsec$ pixels. For each galaxy, the 3 spectra are
    shown on the same (logarithmic) intensity scale, to show the decrease in
    the noise levels between the subsequent steps. The arrows refer to
    specific features in the spectra and are discussed in the
    text. \label{fig:specstack}}   
\end{figure*}

Individual exposures were then combined and cosmic rays were rejected
using a simple sigma-clipping criterion. 
The continuum emission of the galaxy and the night-sky emission lines
were removed by fitting low-order polynomials along the spectral and
spatial axes of the spectra respectively. 

In the final, cleaned spectra, the \Ha line is usually the strongest
line in the outer parts of the galaxies. 
In the central parts however, the 6583.46~\AA\ [N{\sc ii}] line and the
[S{\sc ii}] lines are often stronger, presumably due to underlying stellar
absorption in H$\alpha$.   
Rather than first determining rotation curves for each line separately
and then combining them into one single curve, we have chosen the
reverse order. 
The parts of the spectrum around each of the 5 emission
lines were shifted according to the difference in rest wavelength and 
stacked to create a single spectrum which contains emission from all
lines. 
Each line was roughly weighted according to its relative strength; the
weights $w_i$ are given in table~\ref{table:Haobservations}.    
Before stacking, special care was taken to ensure that the different emission
lines trace similar velocities, but we found no cases with significant
differences. 
This procedure has the added advantage that the signal-to-noise ratio
in the stacked spectrum is higher than in the original one; emission
that was too weak to be detected in each line individually could
sometimes be detected with sufficient significance in the stacked
spectrum.  

In figure~\ref{fig:specstack}, we illustrate the stacking procedure for three
representative galaxies. 
In the spectrum of UGC~624 ({\em left}), the \Ha line is the strongest line
along the entire slit, but adding the other lines leads to a slightly higher
signal-to-noise ratio (see for example the location indicated with arrows 1),
and thus improves the accuracy of the fitted velocities.  
For UGC~4605 ({\em middle}), the improvement is more significant. 
In the original spectrum, \Ha is stronger in the outer parts, but the
6583.46~\AA~[N{\sc ii}]-line is stronger in the centre. 
In the stacked spectrum, velocities can be measured in both regions, as well
as at locations where the signal in the individual lines was too weak to be
fitted (arrows 2). 
For UGC~6786 ({\em right}), the stacking procedure does not work due to a
strong stellar \Ha absorption feature  in the centre (arrow 3). 
In this case, stacking the various lines causes the \Ha absorption feature to
dilute the little emission that is present in the 
6583.46~\AA~[N{\sc ii}]-line, and thus degrades, rather than improves, the
quality of the data. 
In this case, we analysed both lines separately, and combined the resulting
rotation curves afterwards (see also the note in appendix~\ref{app:notes}).  

The final cleaned and stacked spectra are shown in the top middle
panels in the figures in appendix~\ref{app:figures}.  

\subsubsection{derivation of the rotation curves}
From the stacked spectra, the radial velocity of the emitting gas was
determined at each position along the slit by fitting Gaussian
profiles along the wavelength direction. 
Before performing the fits, the spectra were binned in the
spatial direction to $\sim \! 1-2\arcsec$ pixels to increase the
signal-to-noise ratio of the data and to ensure that only one
data point is fitted per resolution element. 
In some cases, parts of the spectra had such low-level emission that
the signal-to-noise ratio was still too low in the binned spectra; for
those regions, larger bin sizes were used.  
Spurious fits or fits with very large errorbars were discarded by
hand. 
The fitted velocities are shown overplotted over the binned spectra
in the top middle panels in the figures in
appendix~\ref{app:figures}. 
They are also overplotted over a major-axis slice through the \HI\
data cube, shown in the top right panels in the same figures. 

The radial velocity curves for the approaching and receding sides were
then folded, using the centre of the optical continuum emission as
central position. 
In two cases, UGC~3205 and 3580, the centres of symmetry of the
emission lines appear shifted with respect to the location of the 
brightest continuum emission, in both cases by approximately one
arcsecond (see the figures in appendix~\ref{app:figures}).
In these two cases, we determined by eye the central position which
gave the largest degree of symmetry in the folded rotation curves. 
In the case of UGC~3580, the offset can easily be explained as a
result of absorption of continuum emission by dust (see
appendix~\ref{app:notes}); for UGC~3205, the origin of the
offset is unknown.  

The systemic velocity was determined by taking, at each radius, the
midpoint of the velocities of the approaching and receding sides and
taking the average of the resulting values. 
This procedure maximizes the symmetry between the approaching and
receding sides of the rotation curves over the full length of the
spectra.   
In most cases, the systemic velocity thus derived is consistent with 
the value found from the \HI\ data (see also
figure~\ref{fig:paramcomp}). 
In some cases, small differences were found; this happened mostly in
galaxies which are kinematically lopsided, where an unambiguous
determination of the systemic velocity is difficult.   
In those cases, we closely inspected the optical spectrum and the \HI\
velocity fields and determined interactively the systemic velocity
which led to the smallest asymmetry in the final, combined optical and
\HI\ rotation curves. 

Finally, the radial velocity relative to the systemic velocity was
calculated for each point and, at radii where emission was detected on
both sides of the galaxy, the weighted average was determined.  
The final rotation curves were subsequently derived by correcting the 
average radial velocity curves for the inclination of the galaxy and
for possible misalignments of the slit with the true major axis; the 
values for the inclination and position angle were taken from the
results of the tilted ring fits to the \HI\ velocity fields, described 
above.  
The resulting rotation velocities are shown with the filled circles in
the bottom right panels in the figures in
appendix~\ref{app:figures}. 

\subsubsection{optical beam smearing and other line-of-sight integration
effects}  
\label{subsubsec:optbeamsmear}
Close inspection of the optical spectra reveals that in many cases,
the rotation curves rise so steeply in the centres of the galaxies
that even in the optical spectra, the gradients are not fully
resolved.  
Thus, although the optical spectra are a major improvement over the 
lower resolution of the \HI\ observations, they suffer from the
optical equivalent of beam smearing as well and the fitted velocities
in the central parts may still not represent the actual rotation
velocities.   

Furthermore, many spectra have line profiles that are broadened even
at positions several arcseconds away from the centres of the galaxies,
where lack of resolution is not expected to play a major role anymore.  
These broadened profiles may be the result of line-of-sight
integration effects through the disks and bulges of the galaxies. 
Again, the simple Gaussians which were fitted to these line profiles will not
recover the true radial velocity at the projected radius and cannot be used for
the final rotation curves.    

We have adopted a method similar to the envelope-tracing (or terminal
velocity) technique \citep{Sancisi79,Sofue96,Garcia-Ruiz02} to correct
the inner points of the optical rotation curves which are affected by
optical beam smearing and/or other line-of-sight integration effects.
We determined by eye the terminal velocities of the affected line
profiles, taking into account the instrumental velocity
resolution. The effect of random motions of the emitting gas clouds is
ignored, as it is generally much smaller than the instrumental
broadening of the profiles ($\sim$10 vs.\ $\sim$50 km/s).
The radial velocities that were thus derived were then processed in
the same way as the results from the Gauss fits to derive the average
inner rotation curve. 
The resulting rotation velocities are shown with the open circles in
the figures in appendix~\ref{app:figures}.  

Although the manually corrected rotation velocities are certainly a
better approximation of the true velocities than the results of
simple Gauss fits to the line profiles, there are many uncertainties,
particularly regarding the detailed 3D distribution of the gas, that
cannot be accounted for with the data used here. 
A more rigorous investigation of the kinematics in the central parts
of the galaxies studied here would require even higher spatial
resolution and preferably a fully 2D velocity field, i.e.\ either
space-based or adaptive-optics assisted integral field spectroscopic
observations. 

\subsection{Final steps}
\label{subsec:finalsteps}
For the final rotation curves, the output from the tilted ring fits to the
\HI\ velocity fields was compared to the derived optical rotation curves and
it was determined which \HI\ data points were affected by beam smearing. 
Central \HI\ data points which lay significantly below the optical
velocities were discarded.  
In almost all cases, the effect of beam smearing was limited to 1 -- 2
\HI\ beam sizes from the centre, and only the inner two or three
points of the \HI\ rotation curves had to be rejected.
Only in highly inclined galaxies, such as UGC~4605 or 8699, do beam
smearing and line-of-sight integration effects play a role at larger
radii; these galaxies were treated individually to ensure that optimal
corrections were applied (see appendix~\ref{app:notes}).  
Outside the regions where the \HI\ observations are affected by beam
smearing, the optical and \HI\ rotation curves generally agree to a
high degree ($\la 10$~km/s). 

The remaining \HI\ data points were then combined with the optical
data to produce the final rotation curves.
Our final curves probe the rotation velocities over 2 -- 3 decades
of radii and enable us to measure small scale variations in the inner parts of 
the optical disks as well as the behaviour in the outer parts of the gas
disks, many optical scale lengths away from the centre.  
 
The combined data points and their corresponding errors can be used,
without further manipulation, to fit detailed mass models and to study
the distribution of luminous and dark matter in the galaxies; this will be 
done in a forthcoming publication.  
For the remainder of this paper, we are interested mainly in the
global properties and shapes of the rotation curves. 
For this purpose, it is helpful to remove the statistical fluctuations between
the individual data points, especially those from the optical spectra.  
To do so, we fitted cubic splines through the data points, using the
interactive fitting task CURFIT in IRAF. 
Individual data points from the rotation curves were weighted
according to their errors (see below); points that were clearly offset 
from the main rotation curve were eliminated during the fits. 
The resulting curves are smooth but still follow the general behaviour
that underlies the individual data points; they will be used in the
remainder of the paper to study the shapes of the rotation curves
and possible correlations with global properties of the galaxies
(section~\ref{sec:rotcurshape}). 
They are plotted as bold lines in the rotation curve panels in
appendix~\ref{app:figures}.    

Finally, a few basic quantities are derived from the rotation curves. 
The rotation curves were classified on the basis of the quality and
reliability of the data. 
Galaxies which are symmetric, show no signs of strong non-circular
motion and have well-defined orientation angles are classed as
category~I. 
This class contains the following galaxies: UGC~2953, 3205, 4605,
6786, 6787, 8699, 9133 and 12043.  
Category II contains galaxies with, for example, mild asymmetries,
bar-induced streaming motions or signs of interactions or tidal
distortions, as well as galaxies for which the orientation angles
could not be constrained as accurately. 
The following galaxies were classed as category~II: UGC~2487 (Seyfert
nucleus), 2916 (interacting, lopsided), 3546 (strong bar and Seyfert
nucleus), 3580 (lopsided), 3993 (inclination angle uncertain),
4458 (possibly tidally disturbed), 5253 (tidally disturbed), 11670
(large bar), 11852 (bar, kinematically disturbed) and 11914
(inclination angle uncertain). 
UGC~624 was classified as category~III, because of the large-scale
asymmetries present in the optical spectrum and particularly in the
\HI\ velocity field. 
The rotation curve of this galaxy is of insufficient quality to be
used for mass modelling.
The classification of the rotation curves is listed in column (11) of  
table~\ref{table:data}. 

We determined by eye the maximum and asymptotic rotation velocities, 
$V_{\mathrm {max}}$ and $V_{\mathrm {asymp}}$ respectively. 
They are listed in table~\ref{table:data} and indicated with
the horizontal arrows in the bottom right panels of the figures in 
appendix~\ref{app:figures}. 
Similarly, the rotation velocity at 2.2 R-band disk scale lengths,
$V_{\mathrm {2.2h}}$ was determined. 

From the velocity at the outermost point of the rotation curve, the
total enclosed mass is calculated as:
\begin{equation}
  M_{\mathrm{enc}} = \frac{V_{\mathrm{out}}^2 R_{\mathrm{out}}^{}}{G}
  = 2.325 \cdot 10^{5} \left( \frac{V_{\mathrm{out}}}{{\mathrm {km/s}}}
  \right)^2 \frac{R_{\mathrm{out}}}{\mathrm{kpc}} \: \msun, 
\label{eq:enclosedmass}
\end{equation}
with $V_{\mathrm{out}}$ the velocity at the last measured point and
$R_{\mathrm{out}}$ the corresponding radius. 
The resulting masses are listed in table~\ref{table:data} as
well. 
In this calculation, it was implicitly assumed that the mass
distribution interior to $R_{\mathrm{out}}$ is spherical. 
If a significant fraction of the total mass is concentrated in a flat
distribution, the value derived here is an upper limit.

\subsection{Rotation curve errors}
\label{subsec:errors} 
Many factors can cause errors in the derived rotation velocities, both
statistical and sys\-te\-ma\-tic. 
For a meaningful interpretation of the results, it is crucial to make
a reliable estimate of all relevant uncertainties and much effort was
therefore put into the identification and quantification of possible
sources of errors.  

We account for three main contributions to the errors in the rotation 
curves. 
The first is simply the measurement error $\Delta V_m$.
For the \HI\ data, this is given by the ROTCUR algorithm, based on the
dispersion around the fitted tilted ring velocities; for the optical data, it
is the fitted error on the profile centre, given by the Gaussian fitting
routine.  
For the manually adapted velocities, the measurement errors were estimated by
eye, based on the shape of the line profiles and the degree to which the data
are degraded by beam smearing and line-of-sight integration effects.   
The contribution $\Delta V_m$ is usually significant only in the
optical rotation curves and in the inner parts of the \HI\ rotation curve,
where only few points are available on the velocity field. 
At larger radii in the \HI\ rotation curves, where each tilted ring
covers many data points in the velocity field, the measurement error
is usually small ($\sim \!\! 1-2\%$). 

The second contribution $\Delta V_{nc}$ comes from kinematical
asymmetries and non-circular motions in the galaxies. 
These were estimated by deriving rotation curves for the approaching
and receding sides of the galaxies separately. 
Additional tilted ring models were fitted to the approaching and
receding sides of the velocity fields and the resulting rotation
curves were combined with the fitted velocities from the corresponding
parts of the optical spectra. 
The resulting rotation curves are shown with the crosses and
plus-signs respectively in the bottom right panels in the figures in 
appendix~\ref{app:figures}.  
The error in the rotation curve $\Delta V_{nc}$ was then estimated as 
one fourth of the difference between the rotation velocities measured
for each side separately \citep[cf.][]{Swatersthesis}. 
With this, rather ad hoc, assumption, the difference between the
rotation velocity for each side separately and the average value 
represents a $2\sigma$ deviation. 
Note that small-scale non-circular motions, or asymmetries
perpendicular to the major-axis, are not accounted for in this estimate.

The first two contributions to the rotation curve errors, $\Delta V_m$
and $\Delta V_{nc}$ were added quadratically, and are shown with the
errorbars in the figures in appendix~\ref{app:figures}. 

The third contribution to the rotation curve errors comes from the
uncertainty in the orientation of the gas disks.
The main contribution comes from the uncertainty $\Delta i$ in the
inclination angle, estimated as in section~\ref{subsec:HIrotcurs}.   
Errors in position angle are usually much smaller, and moreover, only
contribute in second order to the rotation curve errors; in practice, they can
be neglected compared to the uncertainties in inclination. 
The effect of the inclination errors on the rotation curves is derived as
follows.   
The rotation velocities $V_{\mathrm {rot}}$ in the rotation curve can be
written as $V_{\mathrm {rot}} \propto V_{\mathrm {rad}}/\! \sin i$,
where $V_{\mathrm {rad}}$ is the measured radial velocity from either
the optical spectrum or the \HI\ velocity field.  
Thus, an error $\Delta i$ in the inclination leads to an error $\Delta
V_i$ in the rotation velocity of
\begin{equation}
 \Delta V_i = \frac{V_{\mathrm {rot}}}{\tan i} \, \Delta i_{\mathrm {rad}},
 \label{eq:deltaVi}
\end{equation}
where $\Delta i_{\mathrm {rad}}$ is measured in radians.
So, not only is it more difficult to derive the inclination accurately
for near face-on galaxies, the resulting uncertainty in the rotation
velocities due to a given error $\Delta i$  becomes progressively
larger as well.  

The derived errors $\Delta V_i(r)$ are indicated with the shaded regions
in the bottom right panels in the figures in
appendix~\ref{app:figures}; for clarity, they are drawn around
the smoothed rotation curves, rather than around the individual data
points.  
Note that these errors account not only for systematic {\em offsets}
of the rotation curves (due to a global misfit of the inclination), but also
for the effect of undetected or misfitted warps, which would alter the {\em
shape} of the rotation curves.

\section{Parameter comparison}
\label{sec:paramcomp}
A proper derivation of a rotation curve depends crucially on the
assumed orientation parameters and systemic velocity of the galaxy. 
It is instructive to compare the values which we assumed for the
rotation curves here, derived from the tilted ring fits, with those 
obtained from other sources.  
\begin{figure*}
 \centerline{\psfig{figure=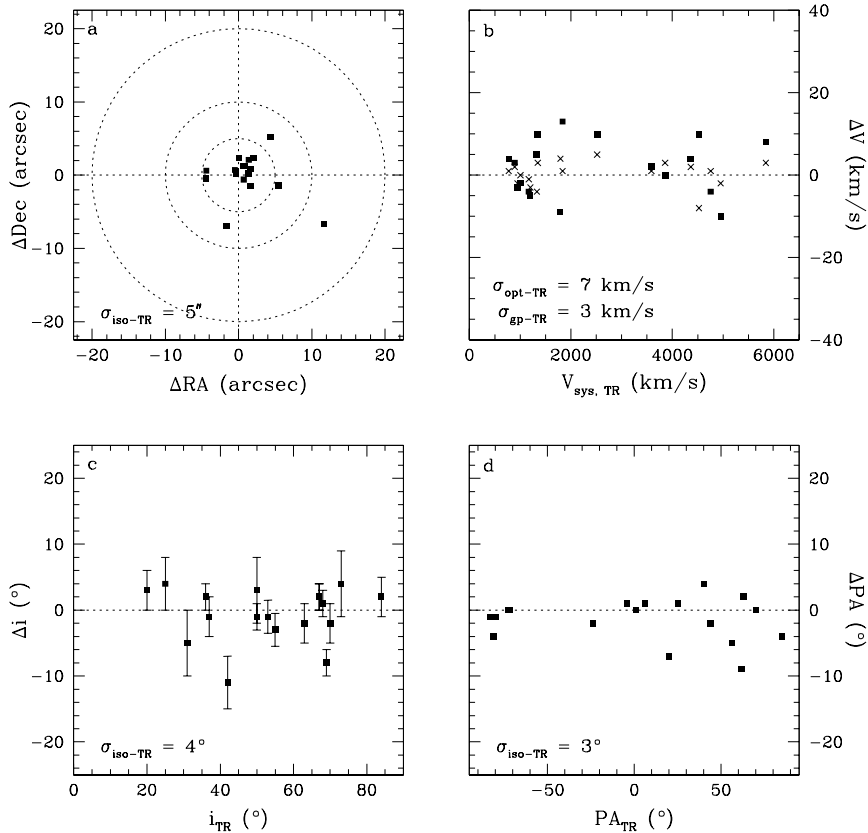,height=11cm}}
  \caption{Comparison of global tilted ring (TR) parameters with values from
    other sources. Data points give offsets of central position (a), systemic
    velocity (b), inclination (c) and position angle (d) with respect to the
    values derived from the tilted ring analysis. For warped galaxies, the
    tilted ring values in the inner regions were used, such that they
    correspond to the same regions as probed by the optical observations. 
    Filled squares represent the values from the optical isophotal analysis
    from \citetalias{Noordermeer06a} (a, c, d) or from the optical spectra
    (b). The crosses in panel b show the comparison with the systemic
    velocities derived from the global \HI\ profiles
    \citepalias{Noordermeer05}. The errorbars in panel c show the adopted
    uncertainties in the tilted ring inclinations. The standard deviations of 
    the distributions are given in the bottom left corner of each panel. Cases 
    where parameters from different sources were not independently derived
    (e.g.\ UGC~624) are not considered here. \label{fig:paramcomp}}  
\end{figure*}

Figure~\ref{fig:paramcomp} shows that, in general, there is
good agreement between the parameters from the tilted ring analysis
and those derived from e.g.\ the optical isophotal analysis. 
In most cases, the dynamical and the isophotal centres coincide within
a few arcseconds, well within one \HI\ beam. 
Larger offsets are only observed in galaxies with strong dust features
in the optical image (e.g.\ UGC~3580) or in highly inclined systems
(e.g.\ UGC~8699), where extinction and line-of-sight integration
effects complicate a proper determination of the central position.  
The observed offsets can be fully explained by observational effects,
and no galaxies seem to have a true, physical offset between the
dynamical and morphological centres.

A comparison of the systemic velocities from different methods shows
that they all agree within 5 -- 10 km/s. 
In galaxies with well-resolved \HI\ velocity fields, however, the
tilted ring systemic velocity is the preferred value, as it uses
dynamical information from the entire gas disk. 
The other methods are expected to have larger intrinsic errors, so the
dispersions given in the figure will probably come predominantly from errors
in those measurements; the typical error with which one can determine the
systemic velocity using tilted ring models is probably of the order of 2 -- 4
km/s.  

The disk orientation parameters derived from the optical isophotes
usually agree within a few degrees with the tilted ring parameters.  
In particular, panel (c) shows that our assumed errors $\Delta i$ on
the inclinations are reasonable. 
Only 2 galaxies show an offset between the isophotal and kinematical
inclination angles that is significantly larger than the assumed
error.
One is UGC~2916, which is interacting with a companion galaxy. 
Close inspection of the fitted optical ellipticities
\citepalias{Noordermeer06a} shows that the shape of the
isophotes at intermediate radii is consistent with the kinematical
inclination; the outer isophotes are most likely disturbed by the
tidal influence of the companion. 
The other case is UGC~6787, where the isophotal inclination is poorly
constrained due to the influence of the dominant bulge. 
Again, since the tilted ring analysis uses dynamical information from
the entire gas disks, it will generally give more accurate values for
the inclination and position angles, so
figure~\ref{fig:paramcomp} mostly shows the uncertainties in
the isophotal parameters. 
Note that the inclinations from LEDA have a large scatter around our
values, with discrepancies up to $20 \deg$.

\section{Warps}
\label{sec:warps}
It has been known for a long time that the outer parts of the gas
disks of many spiral galaxies are not coplanar with the inner disk,
but that they are `warped' \citep{Rogstad74, Sancisi76}. 
\citet{Bosma91} reported that at least 50\% of all galaxies are
warped. 
More recently, \citet{Garcia-Ruiz02} studied 26 edge-on galaxies and
found that {\em all} galaxies with an \HI\ disk more extended than the
stellar one are warped. 

Most galaxies in our sample are fairly face-on and warps are therefore
seen less easily than in Garcia-Ruiz' galaxies.  
Nevertheless, we can infer the presence of warps from the tilted ring
fits to the velocity fields. 
Inspection of the figures in appendix~\ref{app:figures} shows
that the fitted inclination or position angles show significant radial 
variations in 14 of our 19 galaxies.  
For three of the remaining five galaxies (UGC~624, 3993 and 8699), the
quality of the velocity fields is insufficient to put strong
constraints on the orientation of the gas disks and we cannot exclude
the possibility that these systems are warped as well. 
Only two galaxies, UGC~3205 and 3546, show little variation in the fitted
orientation angles (the variations in the inner part of UGC~3205 can be
attributed to bar-induced streaming motions) and seem to have no detectable
warp at all.  

\citet{Briggs90} claimed that warping tends to set in in the outer parts
of the optical disk (around R$_{25}$).  
Although many of the galaxies in our sample are consistent with having
a flat gas disk within the optical radius, we also find a few
counter-examples.  
The velocity fields of UGC~6786, 6787 and 11852 show clear signs of
warping in the inner parts.
The first two systems are unbarred and the observed variations in
their orientation parameters must be real. 
UGC~11852 has a bar, but it is smaller than the \HI\ beam; the
observed warping occurs at larger radii and must, again, be real. 
Note, however, that in all three cases, the inner warps are mild. 
Strong warps are only observed outside the bright optical disks, e.g.\ 
in UGC~9133 and 11852.

\section{Rotation curve shape}
\label{sec:rotcurshape}
The shape and amplitude of a galaxy's rotation curve are directly
related to the gravitational field in the midplane of its disk, and
thus to the mass distribution of its main components. 
A systematic study of the shapes of rotation curves, and a comparison
with the optical properties of the galaxies, can therefore yield
important information on the distribution and relative importance of
dark matter in galaxies. 
In particular, many studies have addressed the correlation between the
distribution of luminous matter and the shape of a rotation curve (see
the introduction for references).
If the luminous matter plays a significant r\^ole in the dynamics of
galaxies, the shape of a rotation curve must depend on the
distribution of the luminous matter. 
On the other hand, if dark matter is dominant, such a correlation will
be much weaker or completely absent. 
\begin{figure*}
 \centerline{\psfig{figure=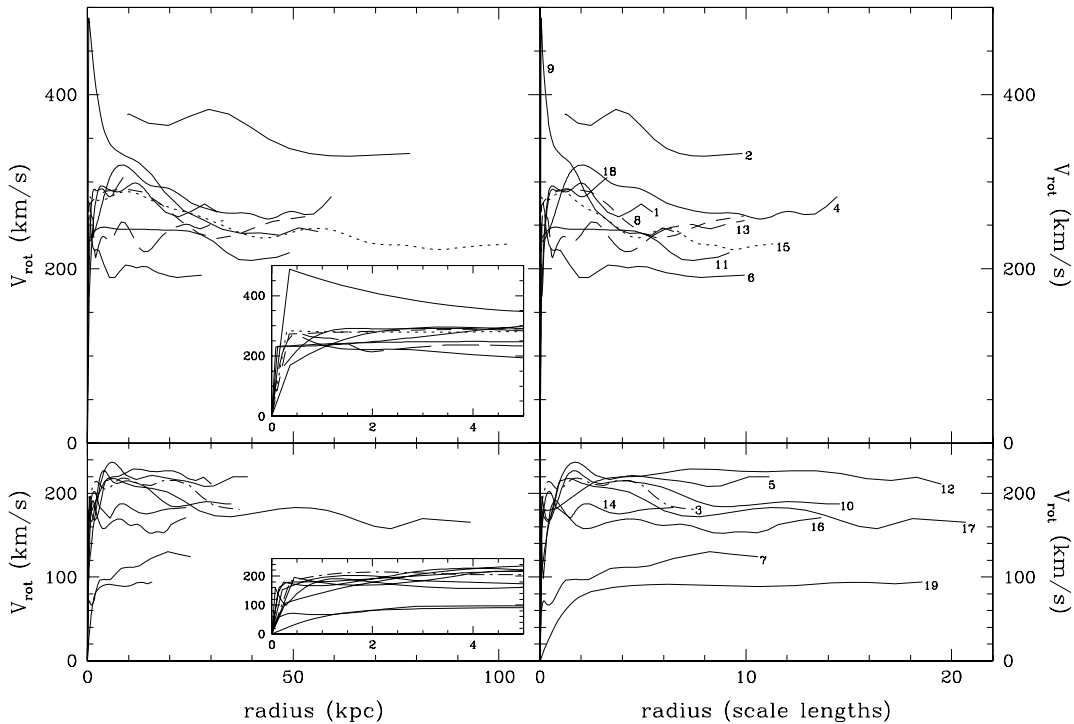,height=9.5cm}}
  \caption{Compilation of rotation curves from all galaxies in the
  sample. The top panels show the rotation curves with $V_{\mathrm
  {max}} > 250 \, {\mathrm {km/s}}$, the bottom panels show those with 
  $V_{\mathrm {max}} < 250 \, {\mathrm {km/s}}$. 
  In the left hand panels, the rotation curves are plotted on
  the same physical scale; the insets show the inner 5~kpc of all 
  curves. In the right hand panels, all radii are scaled with the
  R-band disk scale lengths. The curves are identified with the sample
  numbers from table~\ref{table:sample}. To limit confusion,
  a few curves are plotted using different linestyles: UGC~2916 ({\small \#3},
  dot-dash), 3993 ({\small \#8}, short dash), 6787 ({\small \#13}, long dash)
  and 9133 ({\small \#15}, dots).  
  \label{fig:allrotcurs}}  
\end{figure*}

In figure~\ref{fig:allrotcurs}, we show a compilation of all
the rotation curves in our sample. In the left hand panels, the
rotation curves are plotted on the same physical scale; in the right
hand panels, all radii are scaled with the R-band scale lengths of the
stellar disks \citepalias[from][]{Noordermeer06a}.
Although there is a large variety in rotation curve shape among the
galaxies in our sample, there are also some general features which can
be recognised from this figure and from the individual rotation curves
shown in appendix~\ref{app:figures}. 

Almost all rotation curves in our sample rise extremely steeply in the
central regions. 
In only one case (UGC~12043 ({\small \#19})) do we see the `standard' gradual
solid-body-like rise of the rotation curve, before flattening out at about 3
disk scale lengths. 
In all other cases, the initial rise of the rotation curve is unresolved, even
in the optical spectrum, and the rotation velocities rise from 0 to
$\ga$~200~km/s within a few hundred parsecs (or, similarly, within a fraction
of a disk scale length). 
In some cases (such as UGC~2953 ({\small \#4}), 3205 ({\small \#5}) or
3580 ({\small \#7})), the steep central rise is followed by a more
gentle increase before the maximum rotation velocity is reached; in
other cases (e.g.~UGC~4458 ({\small \#9}), 5253 ({\small \#11}), 9133
({\small \#15})), the rotation curve rises to its maximum immediately.  

At larger radii, many rotation curves show a marked decline.
In several cases (for example UGC~2487 ({\small \#2}), 2916 ({\small 
\#3}), 5253 ({\small \#11})), the rotation curves are more or less
flat in the inner regions and the decline sets in quite suddenly
around the edge of the optical disks (near $R_{25}$); this behaviour
is similar to that in e.g.\ NGC~3992 \citep{Bottema02} and NGC~5055
\citep{Battaglia06}.  
But there are also cases where the  rotation velocities start
decreasing well inside the optical disk (e.g.\ UGC~2953 ({\small
\#4}), 9133 ({\small \#15})), or even right from the first point in the
rotation curve (UGC~4458 ({\small \#9})).    

Although the total decline in the rotation curve can be large (more
than 50\% in the case of UGC~4458 ({\small \#9}); $\sim$~25\% for
UGC~9133 ({\small \#15}) and 11852 ({\small \#17})), all declining
rotation curves appear to flatten out at large radii.  
No rotation curves are found with a fully Keplerian decline in the
outer regions, indicating that we have not yet reached the point where the  
mass density becomes negligible. 
Thus, although the rotation curves of massive, early-type disk
galaxies look remarkably different from those of later-type spiral
galaxies at small and intermediate radii (with the latter generally lacking
the steep rise in the centre and the decline at intermediate radii;
\citealt{Corradi90}, \citealt{Spekkens05}, \citealt{Catinella06}), they show
the same `flatness' in the outer regions, proving that they too must contain
large quantities of dark matter.  
An interesting possible exception is UGC~4458, whose rotation curve only
flattens out in the very outer regions. 
Given the large uncertainties in the outer data points due to the face-on
orientation, we cannot strictly rule out that this rotation curve keeps
declining in Keplerian fashion.  
We will investigate this issue, and its implications for the dark matter
content in this galaxy, in more detail in our subsequent paper on the
mass modelling.  

It is worth mentioning that many galaxies show distinct features in
their rotation curves (e.g.\ UGC~6787 ({\small \#13}), 8699 ({\small
\#14}); see also the notes on individual cases in
appendix~\ref{app:notes}).  
Only few galaxies have smooth rotation curves without `bumps' or `wiggles' and 
the declines at intermediate radii are rarely featureless and monotonous. 
Although these irregularities may sometimes be caused by e.g.\ noise
or non-circular motions of the gas, they can often be recognized on
both sides of the optical spectra or \HI\ velocity fields and must, in
most cases, reflect small-scale features in the underlying mass
distribution.  
In particular, we will show in our forthcoming publication on the mass models
that the `wiggles' and the detailed shape of the drop-off in the rotation
curves can, in some cases, be linked to features in the light or gas 
distributions and can be used to constrain the relative contributions
of the luminous and dark matter in these galaxies. 
\begin{figure*}
 \centerline{\psfig{figure=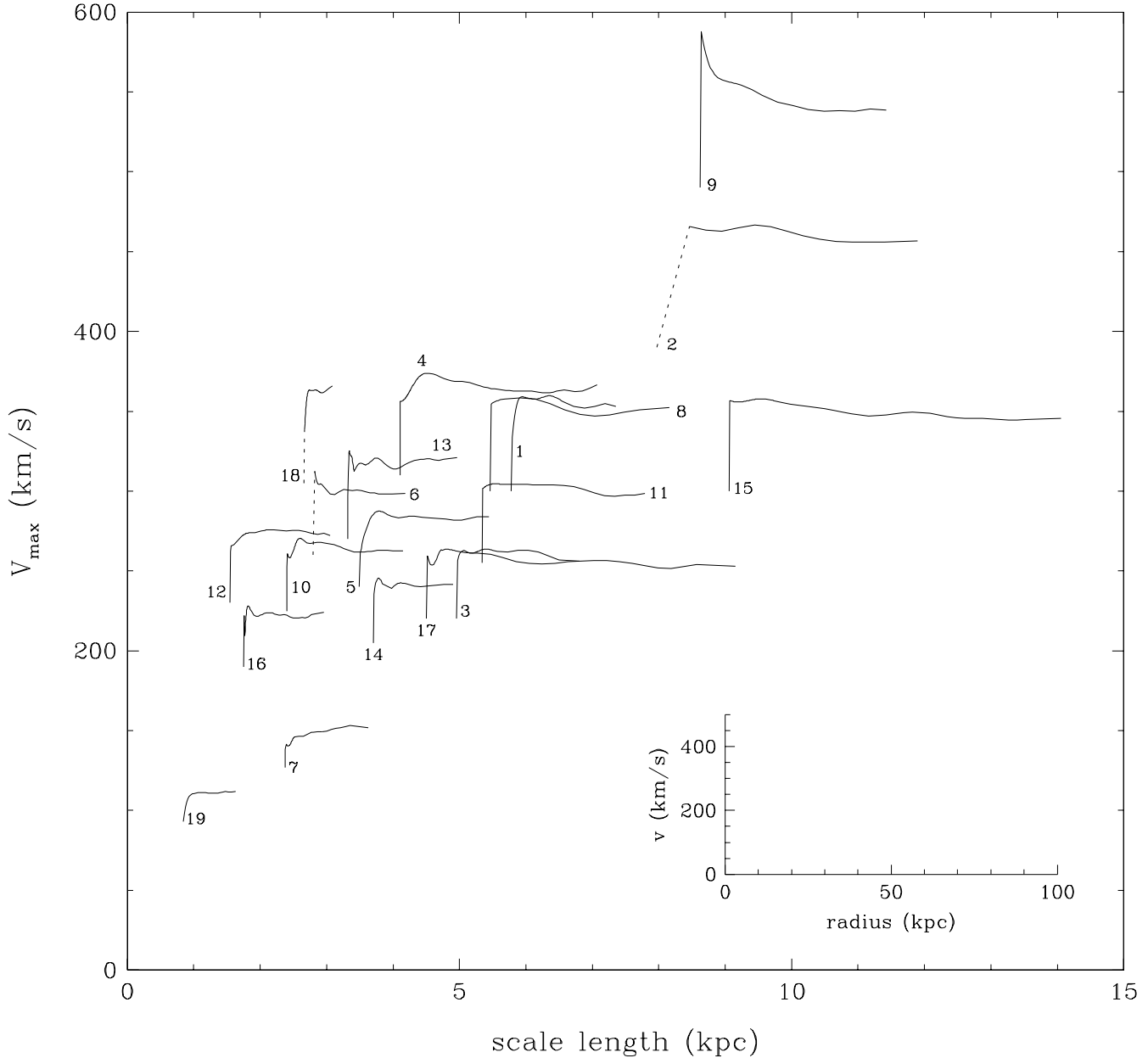,height=16.5cm}}
  \caption{Compilation of rotation curves following \citet{Casertano91}.  
  The origin of each rotation curve is placed according to the maximum
  rotation velocity $V_{\mathrm {max}}$ and the R-band exponential
  scale length of the stellar disk. 
  Dashed lines are used to indicate the origins for galaxies where the
  central rotation curve was not measured. 
  The individual rotation curves are labelled using the sample numbers from
  table~\ref{table:sample} and scaled in radius and velocity,
  as indicated with the small axes at the bottom right. 
  The scale lengths for UGC~4605 and 6786 (\#10 and \#12) are
  estimates only, so their exact position in the figure is uncertain.
  \label{fig:casertanogram}}  
\end{figure*}

In some cases, such as UGC~2953 ({\small \#4}), 3993 ({\small \#8}) or
11670 ({\small \#16}), there are indications that the rotation curves
start to rise again at the outer edges of the \HI\
disks. \label{risingrotcurs} 
Whether this effect is real or an artefact in the data is hard to 
tell.  
The corresponding points in the \HI\ velocity fields were derived from 
low signal-to-noise ratio line profiles and have large uncertainties. 
Furthermore, we cannot exclude the possibility that the gas in the
outer regions moves on non-circular orbits, or that we have not
determined the inclination of the orbits correctly.
Follow-up observations at higher sensitivity are required to
investigate this in more detail.

\subsection{Correlations with optical properties}
\label{subsec:optcorr}
\begin{figure*}
 \centerline{\psfig{figure=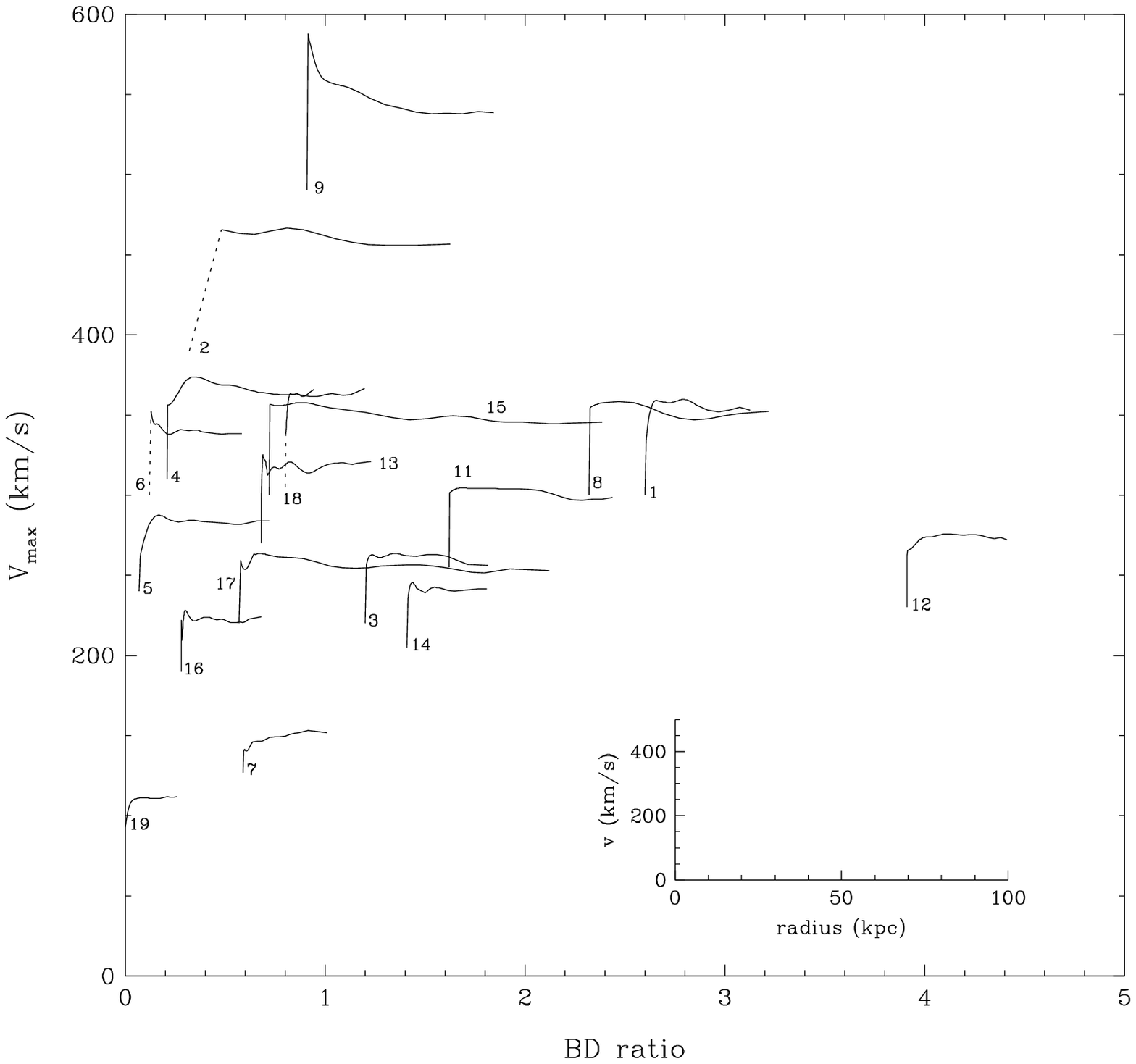,height=16.5cm}}
  \caption{Compilation of rotation curves similar to
  figure~\ref{fig:casertanogram}, this time with the origin of
  each curve placed according to the maximum rotation velocity
  $V_{\mathrm {max}}$ and the R-band bulge-to-disk luminosity ratio. 
  UGC~4605 was not included in this figure, since no bulge-disk
  decomposition was available for this galaxy. 
  \label{fig:BDcasertanogram}}  
\end{figure*}
\begin{figure*}
 \centerline{\psfig{figure=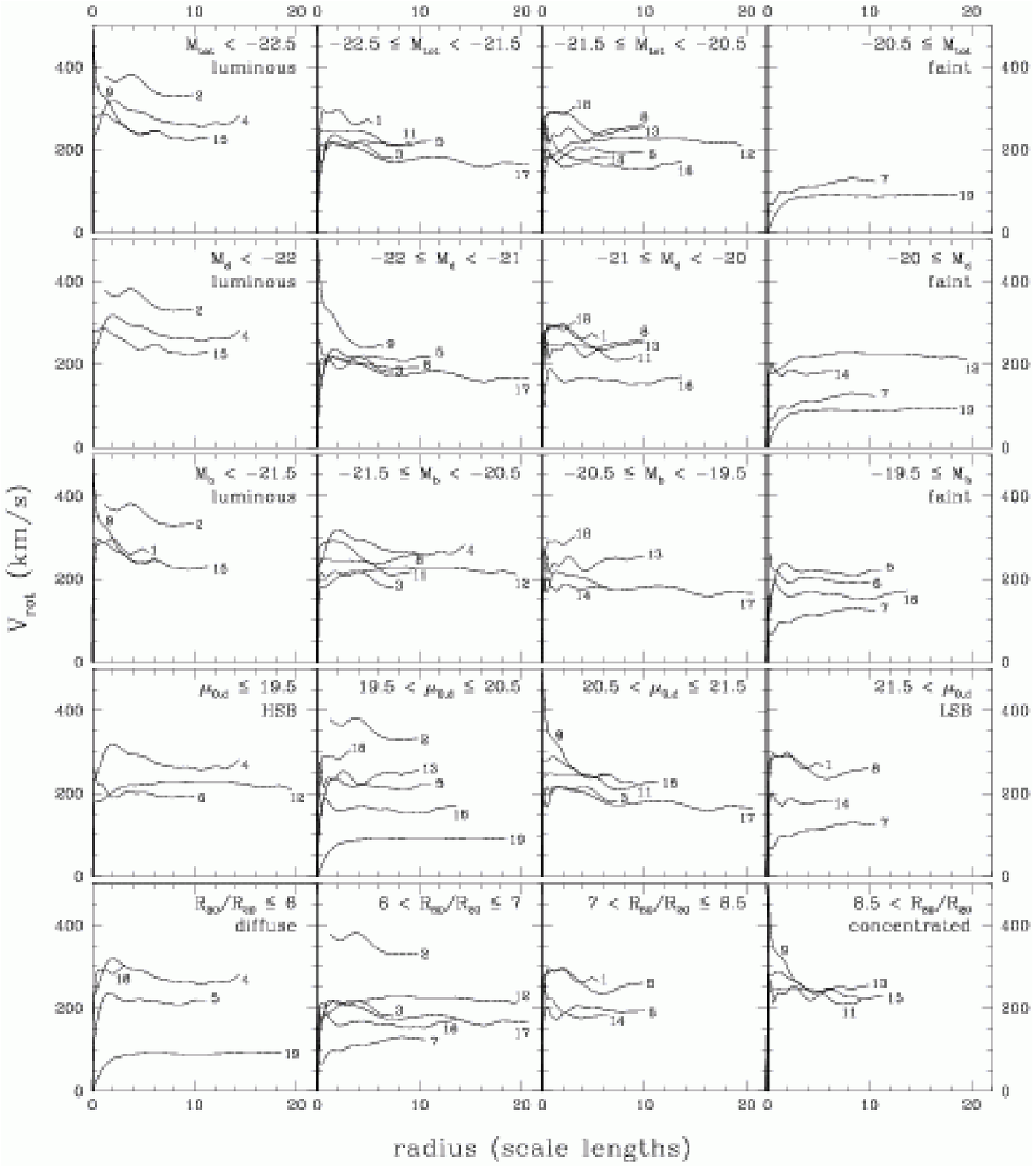,height=17cm}}
  \caption{Rotation curves ordered according to (from top to bottom) 
  absolute magnitude of the entire galaxy, absolute magnitude of the
  stellar disk, absolute magnitude of the stellar bulge, central  
  surface brightness of the stellar disk and compactness of the 
  stellar light distribution (measured by the ratio of the effective 
  radii $R_{80}$ and $R_{20}$ which contain respectively 80 and 20\%
  of the light). All parameters are derived from the R-band images
  \citepalias[see tables~A3 and A4 in][]{Noordermeer06a}. The bins are chosen
  in order to distribute the rotation curves evenly over the panels. All
  rotation curves are labelled with the sample numbers from
  table~\ref{table:sample} and scaled with the R-band disk
  scale lengths. Since no accurate photometry is available for
  UGC~4605 (\#10), this galaxy is not included in these plots. 
  Furthermore, UGC~12043 (\#19) does not have a bulge component and is
  not included in any of the plots in the third row. 
  \label{fig:binnedrotcurs}}  
\end{figure*}

To investigate the dependence of rotation curve shape on the optical
properties of the galaxies, we have ordered the rotation curves from 
our sample according to several parameters. 
In figures~\ref{fig:casertanogram}
and~\ref{fig:BDcasertanogram}, we present a compilation  
of our rotation curves in a similar fashion as \citet{Casertano91};
the rotation curves are ordered according to the maximum rotation
velocity $V_{\mathrm {max}}$ and the R-band disk scale length
(figure~\ref{fig:casertanogram}) or bulge-to-disk luminosity
ratio (figure~\ref{fig:BDcasertanogram}).  
In figure~\ref{fig:binnedrotcurs}, we have divided our
galaxies into different subsamples, according to several optical  
parameters, and plot the rotation curves for each subsample
separately. 

\subsubsection{inner rotation curves}
Early results by \citet{Rubin85} showed that the inner shape of a
rotation curve is coupled to a galaxy's luminosity: bright galaxies
have steeply rising rotation curves, whereas low-luminosity systems
reach the maximum rotation velocity at relatively larger radii.
This relation was later confirmed by several other studies
\citep[e.g.][]{Broeils92, Persic96, Verheijen97, Swatersthesis}.
The galaxies with the lowest luminosity (and corresponding maximum
velocity) in our sample (UGC~3580 ({\small \#7}) and 12043 ({\small
\#19})) follow this trend and have rotation curves which rise
relatively slowly.
In particular, UGC~12043 ({\small \#19}) is the only galaxy in our
sample which completely lacks the characteristic steep rise in the
centre; instead, its rotation velocities increase gradually, in
solid-body fashion and only reach the maximum around 3~disk scale
lengths.  
The remaining galaxies in our sample, however, seem to indicate that
the systematic progression breaks down above a maximum rotation
velocity of $\sim \,$200~km/s (see
figure~\ref{fig:casertanogram}).   
All galaxies with a rotation velocity larger than $\sim \,$200~km/s
have the characteristic steep rotation curve in the centre.  
Whether the rotation velocities continue to increase after this
initial rise, or whether the maximum is reached in the very centre,
does not seem to depend on the total luminosity of the galaxy.  

Instead, the shape of the rotation curve in the inner regions seems to 
depend more strongly on the concentration of the stellar light distribution. 
This can be seen most clearly in figure~\ref{fig:BDcasertanogram} and in the
bottom panels of figure~\ref{fig:binnedrotcurs}, where the rotation curves are
ordered according to the bulge-to-disk luminosity ratio and the more generic
measure of light concentration $R_{80}/R_{20}$ respectively. 
These figures show that the rotation curves of galaxies with faint bulges and
a relatively diffuse stellar light distribution continue to rise after the
steep central part, and reach the maximum outside the bulge-dominated regions
(e.g.\ UGC~2953 ({\small \#4}) and 3205 ({\small \#5})). 
On the other hand, the rotation curves of galaxies with highly concentrated
light distributions rise to the maximum immediately. 
This also explains why UGC~12043 ({\small \#19}) has such a shallow
central rotation curve: it has no bulge component at all. 
The only system with a small bulge which appears to reach its maximum
rotation velocity at very small radii is UGC~3546 ({\small \#6}), but
this galaxy has a Seyfert nucleus which makes its central rotation
velocities highly uncertain (see the errorbars in the figure in
appendix~\ref{app:figures}); it is well possible that the rotation curve of
this galaxy rises more slowly than we have derived here.  
 
Thus, our data appear at odds with the claim of \citet{Rubin85} and
\citet{Burstein85} that optical morphology does not influence the 
shape of a rotation curve and that large amounts of dark matter must be
present at all radii. 
Our data indicate that at least the bulge stars have a strong
influence on the central rotation curves, and suggest that they
dominate the gravitational potential in the inner regions (in
agreement with \citealt{Corradi90}, \citealt{Verheijen97} and
\citealt{Sancisi04}).      

\subsubsection{outer rotation curves}
\begin{figure*}
 \centerline{\psfig{figure=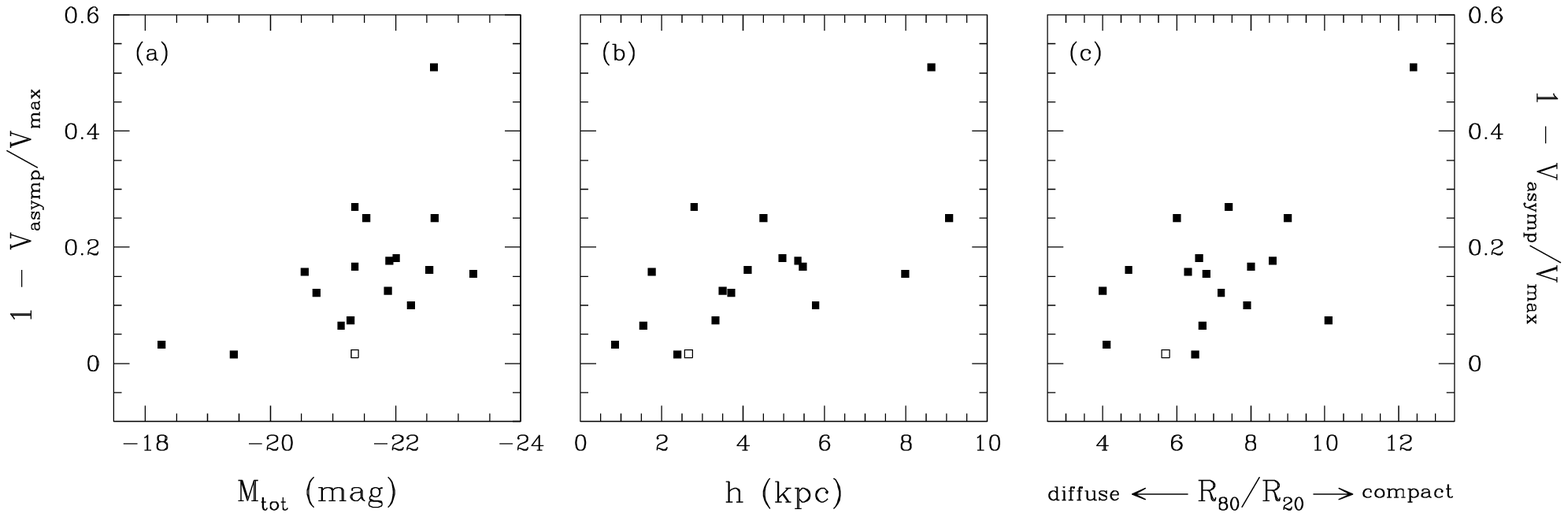,height=4.8cm}}
  \caption{Strength of the decline in the rotation curves vs. total
  magnitude (a), scale length of the stellar disk (b) and ratio of the 
  effective radii $R_{80}$ and $R_{20}$ (c) (all measured from the
  R-band images). The open symbol indicates UGC~11914; due to the
  small radius of the last measured point in the rotation curve, its 
  asymptotic rotation velocity is poorly defined (see also
  appendix~\ref{app:notes}). Since no accurate photometry is
  available for UGC~4605, this galaxy is not included in these plots.   
  \label{fig:decline}}  
\end{figure*}
Inspection of figures~\ref{fig:casertanogram} --
\ref{fig:binnedrotcurs} shows that the shape of the rotation
curves in the outer parts is correlated with the luminosity of the
galaxies: luminous galaxies are more likely to have a declining
rotation curve than low-luminosity systems (in agreement with
\citealt{Casertano91} and \citealt{Broeils92}).  
This is shown in a more quantitative way in panel a) of
figure~\ref{fig:decline}, where we plot the ratio of
asymptotic and maximum rotation velocity as a function of total
absolute magnitude.  
This figure shows that {\em all} early-type disk galaxies with $M_R < -20$
have at least a modest decline in their rotation curve. 

The strength of the decline shows, however, little dependence on
the {\em shape} of the light distribution. 
\citet{Casertano91} concluded that, in a sample of galaxies of type Sb and
later, the most strongly declining rotation curves occur in systems with a
compact light distribution, where `compact' in their terminology meant `small
disk scale length'. 
Our data show that such a correlation does not exist for early-type disks, as
our sample also contains a number of galaxies with large scale lengths which
have falling rotation curves (see figure~\ref{fig:casertanogram}). 
This is in agreement with \citet{Broeils92}, who also found a number
of large galaxies with declining rotation curves.
In fact, two of the galaxies in our sample with the most strongly
declining rotation curves (UGC~4458 ({\small \#9}) and 9133 ({\small
\#15})), have large scale lengths (8.6 and 9.1~kpc respectively).  
Panel b) in figure~\ref{fig:decline} shows that, if a trend
with linear size of the galaxies exists at all, it is in the opposite
direction as observed by \citet{Casertano91}: larger galaxies have on
average more strongly declining rotation curves.  

No trend is seen when, instead of the disk scale length, we use the more
generic parameter $R_{80}/R_{20}$ to define the compactness of the stellar
light distribution (bottom panels in figure~\ref{fig:binnedrotcurs} and panel
c) in figure~\ref{fig:decline}). 
Declining rotation curves are seen both in galaxies with a compact
light distribution and in galaxies with a more diffuse stellar
component (such as UGC~2487 ({\small \#2}) or 2953 ({\small \#4})).  
Note, however, that according to this criterion, the galaxy with the
strongest decline in its rotation curve is also the most concentrated:
UGC~4458 ({\small \#9}).

\subsection{The Universal Rotation Curve for early-type disk galaxies}
\label{subsec:URC}
\citet{Persic96} claimed, based on a study of over 600 optical rotation curves
and a small number of \HI\ rotation curves, that the shape of a rotation curve
is solely governed by the galaxy's luminosity and can be described by the
following simple formula: 
\begin{eqnarray}
V_{\mathrm {URC}}(x) & = & V_{\mathrm {opt}} 
             \left\{ (0.72 + 0.44 \log \lambda) 
                     \frac{1.97 x^{1.22}}{(x^2 + 0.78^2)^{1.43}} 
                   \right. \nonumber \\ 
           &   & \left. \;\;\;\;\;\;\;\;\;\; + 1.6 \, e^{-0.4 \lambda}
                     \frac{x^2}{x^2 + 1.5^2 \lambda^{0.4}}
            \right\}^{1/2}.
\label{eq:URC}
\end{eqnarray}
Here, $x = R/R_{\mathrm {opt}}$ is the radius expressed in units of the
optical radius $R_{\mathrm {opt}}$, the radius encompassing 83\% of the
light.
$V_{\mathrm {opt}}$ is the rotation velocity at $R_{\mathrm {opt}}$ and
$\lambda = L_B/L_B^*$ is the B-band luminosity of the galaxy scaled with
$L^*$.    
In principle, $V_{\mathrm {opt}}$ can also be related to the luminosity via
the Tully-Fisher relation, but since we are mostly interested in the {\em
shape} of the rotation curve here, we empirically determine $V_{\mathrm
{opt}}$ from our observed rotation curves. 
\begin{figure*}
 \centerline{\psfig{figure=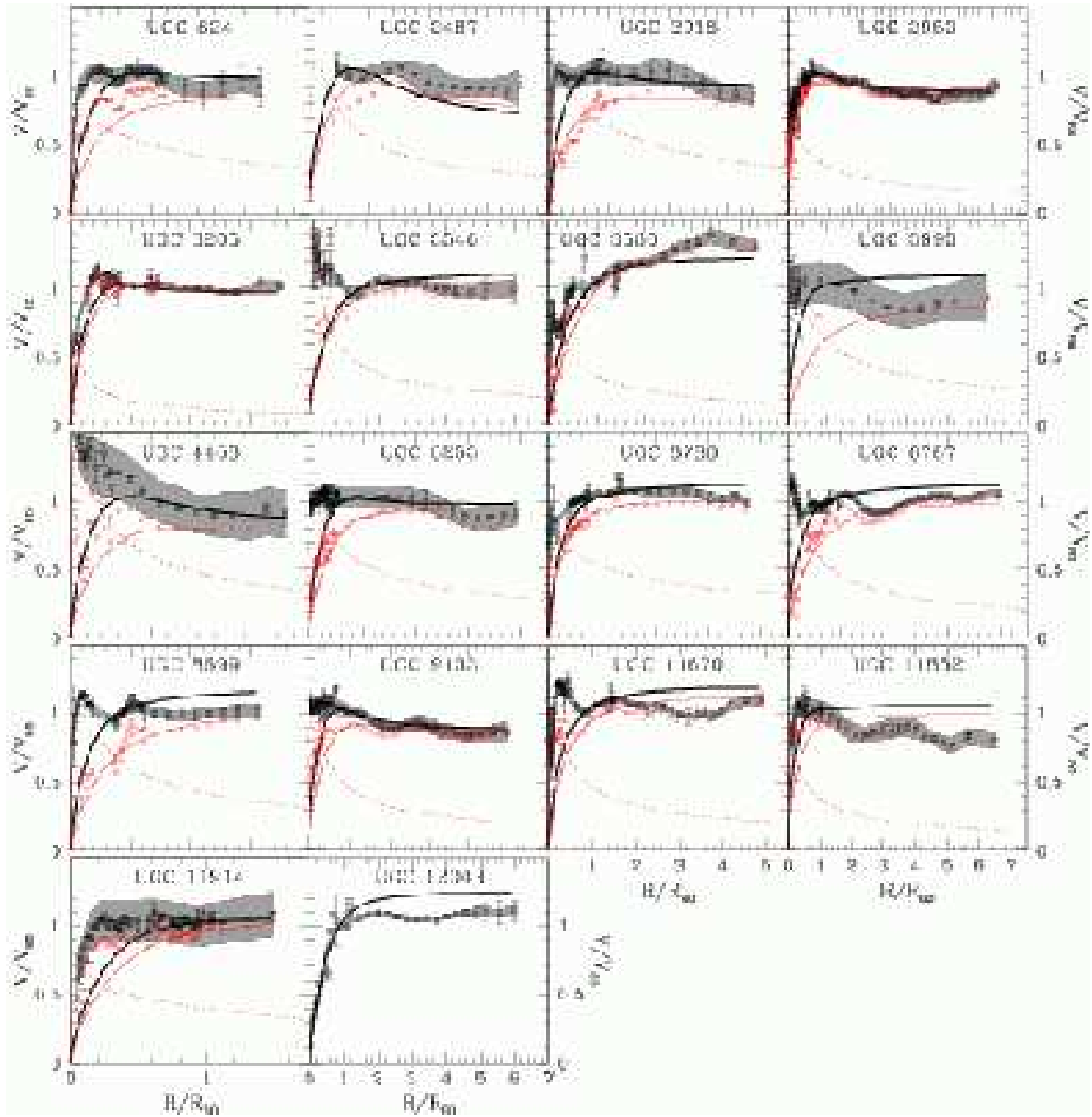,height=17.cm}}
  \caption{Comparison of our rotation curves (black data points with error
  bars) with the Universal Rotation Curve from \citet{Persic96} (bold black
  lines). Shaded regions give the uncertainties due to inclination errors. Red
  data points show the rotation curves after subtraction of the bulge
  component (shown with the dashed red lines; not for the bulgeless galaxy
  UGC~12043). The solid red line shows the URC for the disk component. All
  rotation curves are scaled with $R_{80}$, the radius containing 80\% of the
  total B-band light \citepalias[from][]{Noordermeer06a} and $V_{80}$, the
  rotation velocity at $R_{80}$. See text for details.
  \label{fig:URC}}  
\end{figure*}

This Universal Rotation Curve (URC) has received considerable
attention in the literature, as it implies (together with other
scaling relations such as the Tully-Fisher relation) a tight connection
between the luminous and dark matter in galaxies and, as such, has
important consequences for the theory of galaxy formation
\citep{Dalcanton97, Hernandez98, Elizondo99}.
However, from the observational point of view, no consensus has yet
been reached concerning the general applicability of the concept of
the URC to real galaxies.  
Although the URC seems to give a reasonable description of the general 
trends in rotation curve shapes, it was readily noted that individual
rotation curves often show large deviations from the URC
\citep{Courteau97, Verheijen97, Willick99, Garrido04} and that other
parameters than luminosity must also influence a galaxy's rotation curve
(e.g.\ surface density, bulge-to-disk ratios, etc., see also
\citealt{Roscoe99}).   
Our findings that the rotation curves of early-type disk galaxies have
distinctly different shapes (steep central rise, decline at intermediate
radii) than those of later-type systems of similar luminosity, and that within
our sample, the rotation curve shape is only weakly coupled to luminosity and
rather depends on factors such as light concentration, bulge-disk ratio, etc.,
raise additional questions on the ability of the URC to describe rotation
curves of all classes of disk galaxies.  

In figure~\ref{fig:URC}, we compare the predicted rotation curves from
equation~\ref{eq:URC} to our observed rotation curves.
We did not measure the optical radius $R_{\mathrm {opt}}$ for our galaxies,
but use $R_{80}$, the radius containing 80\% of the light in the B-band
\citepalias[see table~A3 in][]{Noordermeer06a}, as an approximation instead.  
All velocities are scaled with $V_{80}$, the rotation velocity at $R_{80}$.
UGC~4605 was omitted from the analysis, since no accurate photometric data
were available.  
The most obvious result from figure~\ref{fig:URC} is that the
URC completely fails to account for the steep central rise in our
rotation curves.  
In all galaxies, except the bulgeless system UGC~12043, the URC
severely under-predicts the rotation velocities in the centre.
For UGC~2953, it is hard to see the discrepancy in the figure due to
the crowding of the points, but also in this case, the observed
velocities inside $0.4 \, R_{80}$ lie far above the predicted curve. 
This failure in the inner regions is, however, not surprising, as
\citet{Persic96} derived their relations specifically for
disk-dominated galaxies and did not take bulges into account.

In the outer parts, the agreement is good in some cases
(e.g.\ UGC~2953, 3205, 9133), but there are also many galaxies where
the observed rotation curves have a markedly different shape than the
predictions from the URC.
In particular, the slope of the outer rotation curve seems to be
poorly predicted by the URC, with the observed rotation curves flat
instead of rising (e.g.\ UGC~6786, 11670, 12043), declining instead of
flat (UGC~11852) or not declining rapidly enough (UGC~2487).  
But some of these differences at larger radii may be related to the
presence of the bulges as well. 
Not only do bulges influence the observed rotation velocities in the
galaxies, they also change the shape of the predicted rotation curve
by altering the total luminosity and the optical radius; thus, they
may cause discrepancies over the full length of the rotation curves. 

To investigate to what extent the difference between the observed rotation
curves and the URC can be explained by the presence of the massive bulges in
our galaxies, we have subtracted their contribution from the observed and
predicted rotation curves; the results are shown in red in
figure~\ref{fig:URC}.   
The bulge contributions were taken from our mass models, which will be
presented in a forthcoming publication. 
The solid red lines show the predicted rotation curves from
equation~\ref{eq:URC}, now using the absolute B-band magnitudes of the disk
components \citepalias[taken from][]{Noordermeer06a}. 

The correction for the bulge influence has indeed alleviated some of the
discrepancies, especially in the central parts; some galaxies show almost
perfect agreement with the predicted rotation curves now (e.g.\ UGC~2953,
6786, 9133).   
However, even after the bulge corrections, many differences remain.  
Most rotation curves still rise more rapidly than predicted by the URC
and reach a flat plateau where the URC is still rising (e.g.\ UGC~624,
3205, 3546, 3993, 4458, 11670, 11852). 
Also in the outer regions, the observed slopes of the rotation curves
often still differ from the predicted ones, although the differences
are generally smaller than in the original curves. 

In conclusion, the foregoing analysis confirms the dependence of rotation
curve shape on morphological type, with bulge-dominated, early-type disk
galaxies having distinctly different rotation curves than late-type spirals of
similar luminosity. 
Thus, a universal rotation curve that depends only on luminosity is inadequate
to account for the observed diversity in rotation curve shapes along the
Hubble-sequence.
Although the URC of \citet{Persic96} may have its virtue as an empirical tool
to parameterise global trends of several properties of {\em disk-dominated}
galaxies (surface brightness, size, etc.) with luminosity and the reflection
of those on the gravitational fields, additional parameters are required to
account for the presence of bulges in earlier-type disks. 
In addition to the deviations in the inner regions, we have shown that the
detailed shape of the drop-off in the outer parts of our rotation curves is
not well reproduced either.   
Clearly, real galaxies are more complex than the simple URC prescription
suggests and other factors than luminosity must contribute to the detailed
shape of a rotation curve as well.

\section{Discussion and conclusions}
\label{sec:conclusions}
In this paper, we have derived rotation curves for a sample of
19~early-type disk galaxies (S0$^-$ -- Sab) spanning almost 2 decades in
optical luminosity. 
The majority of the galaxies are luminous, with $M_B < -20$. 
The rotation curves were derived from a combination of \HI\ synthesis
observations and long-slit optical spectroscopy of the ionised gas and probe 
the rotational velocities and mass distributions on scales ranging from
100~pc to 100~kpc. 
Almost all of the rotation curves share a number of properties, which appear 
to be typical for this type of galaxies. 

The rotation velocities generally rise rapidly and often reach values of 200
-- 300~km/s (and up to 500~km/s for extreme cases such as UGC~4458) within a
few hundred parsecs from the centres of the galaxies.   
After the initial steep rise, the rotation curves show a diversity in
shapes. 
In some cases, the rotation velocities gradually increase further and
reach the maximum at intermediate radii. 
In other cases, the rotation curves remain flat after the initial
rise, or even start to decline immediately. 
This diversity in shape appears to be related to differences in the light
distribution in these galaxies: galaxies with concentrated light distributions
and luminous bulges generally reach the maximum rotation velocity at small
radii, whereas galaxies with a more diffuse stellar component, or small
bulges, generally have rotation curves which peak further out. 

At larger radii, most rotation curves decline, with the asymptotic rotation
velocity typically 10~--~20\% lower than the maximum. 
The strength of the decline is coupled to the luminosity of the galaxy, more
luminous galaxies having on average more strongly declining rotation curves,
in agreement with \citet{Casertano91}.  
However, we cannot confirm another claim of \citeauthor{Casertano91}, that
declining rotation curves occur preferentially in galaxies with a compact
light distribution. 
By `compact', these authors meant `small disk scale length'. 
In agreement with \citet{Broeils92}, our sample also contains a number of
galaxies with large scale lengths which have falling rotation curves. 
Interestingly, two recent studies \citep{Spekkens05, Catinella06} showed that
later-type galaxies, even those with high optical luminosity, do {\em in
general} not have declining rotation curves.  
In contrast, we find that declining rotation curves are a {\em characteristic
feature} of massive, {\em early-type} disk galaxies. 
This seems to suggest that, although \citeauthor{Casertano91} were correct to
claim that the shape of the light distribution determines whether or not a
galaxy has a declining rotation curve, the term `compactness' must be
interpreted as the presence of a light concentration (i.e.\ a bulge) in the
centre, rather than a small scale in absolute terms. 
Note that the two galaxies in \citeauthor{Casertano91} with the most strongly
declining rotation curves (NGC~2683 and NGC~3521) both have a sizable bulge
\citep{Kent85}.
Within in our own sample, on the other hand, the strength of the decline in
the rotation curves seems barely related to the bulge-to-disk luminosity
ratio, or generic light concentration $R_{80}/R_{20}$.  
Thus, although our data suggest that the bulge plays an important r\^ole in
this issue, we conclude that it is not possible to extract a single parameter
from the light distribution of a galaxy which uniquely determines whether or
not it has a declining rotation curve.
Rather, it must depend in a more subtle manner on the relative masses, and the
details of the mass distributions, of the various luminous and non-luminous
components in a galaxy.  
We will address this issue in more detail in our forthcoming publication on
the mass models for our galaxies.

It is important to note that we have not found any rotation curve which
declines in Keplerian fashion. 
In fact, all rotation curves flatten out in the outer regions.
Early-type disk galaxies, despite appearing dominated by luminous matter in
the central parts, must also contain large amounts of dark matter to explain
the shape of the rotation curves in the outer regions.

Two low-luminosity galaxies, UGC~3580 and 12043, have a distinctly
different kinematical structure than the other systems in our sample.    
The rotation curve of the former does rise rapidly in the centre,
but where in most other galaxies the rotation velocities decrease or
remain constant at large radii, they continue to rise almost all the
way till the outer point in the case of UGC~3580 (out to $\sim \! 8.5$ 
disk scale lengths).  
UGC~12043 completely lacks the steep central rise in the rotation
curve; instead, its rotation velocities increase gradually, in
solid-body fashion, before becoming constant outside approximately 3
disk scale lengths. 
The rotation curves of these two galaxies resemble those of typical
late-type and dwarf galaxies which generally have slowly rising
rotation curves too \citep{Broeils92,Swatersthesis}.
UGC~3580 and 12043 also have different optical morphologies than most 
galaxies in our sample (see \citetalias{Noordermeer06a}) and the
conclusion seems justified that low-luminosity early-type disks form an
entirely different class of galaxies. 

We have compared our rotation curves with the predictions from the
Universal Rotation Curve from \citet{Persic96}. 
Since their model contains only one parameter, the total luminosity, and does
not include morphological type, their URC fails to account for the steep
central rise in our observed rotation curves.  
These discrepancies are reduced, but not removed entirely, when we
subtract the bulge influence from the rotation curves. 
Furthermore, there are also many differences between the observed and
predicted rotation curves at larger radii. 
The concept of a Universal Rotation Curve which depends only on luminosity
appears to be insufficient to account for the observed diversity in rotation
curve shape; other factors must contribute to the detailed shape of a rotation
curve as well.  

All in all, the results presented here show that rotation curves form a
multi-parameter family. 
Although luminosity is clearly a major factor determining the shape of a
rotation curve, other parameters are important too. 
In particular, early-type disk galaxies have distinctly different rotation
curves than their later-type counterparts, an effect which we have shown is
mostly due to the presence of bulges in these systems.  
This is in contrast with some previous claims \citep[e.g.][]{Rubin85,
Burstein85, Persic96} that the shape of the rotation curves is determined by a 
galaxy's luminosity only and that the way the light is distributed has little
influence.
Our results contradict this and, in fact, indicate that above rotation 
velocities of about 200~km/s, the total luminosity has little impact on the
shape of the central rotation curve and that, instead, the {\em shape} of the
stellar light distribution governs the dynamics in the inner parts.
Our findings have important consequences for our understanding of the
structure of galaxies. 
In particular, our data strongly suggest that, at least in the central regions
of the early-type galaxies presented in this study, the luminous matter
dominates the gravitational potential, with dark matter only starting to play
a role outside the bulge-dominated regions.   
We will investigate this issue in more detail in a forthcoming publication,
where we construct detailed mass models for the galaxies in our sample and
study the relation between luminous and dark matter in a more quantitative
way.

\section*{Acknowledgements}
We would like to thank Benne Holwerda for kindly providing the optical spectra
of UGC~2953. We are grateful to Jacqueline van Gorkom and Reynier Peletier for 
stimulating discussions which helped to improve the early stages of this
paper. We would also like to thank the anonymous referee for pointing out
several unclarities in the original document, and for helpful suggestions to
improve the presentation.

\bibliographystyle{mn2e}
\bibliography{../../references/abbrev,../../references/refs}

\clearpage

\appendix


\section{Notes on individual galaxies}
\label{app:notes}
{\bf UGC~624} (NGC~338) has strongly lopsided kinematics, which makes
it difficult to determine the systemic velocity accurately. 
When the systemic velocity is left as a free parameter in the tilted
ring fits, it shows a gradual decline of almost 50~km/s towards larger
radii, in an attempt to symmetrize the rotation curve. 
The value of 4789~km/s minimizes the asymmetries in the central parts,
but even then there are differences between the approaching and
receding sides of the optical spectrum. 
On the approaching side, the rotation curve rises rapidly to a
more or less flat plateau at about 285~km/s, whereas on the receding
side, the rotation curve rises more gradually to a peak of
approximately 310~km/s, after which it declines slowly. 
In the outer parts of the \HI\ disk (R $\ga$ 50~kpc), the differences
are even more pronounced. 
On the receding side, the rotation curve declines gradually, whereas
it starts to rise again on the approaching side. 
At the outermost point, the difference between the two halves amounts to 
almost 100~km/s. \\
It seems plausible that the asymmetries in UGC~624 are caused by
gravitational interaction with its neighbour, UGC~623.  
Note that the distribution of the neutral gas in UGC~624 is also
asymmetric (cf.\ \citetalias{Noordermeer05}). \\
The strong asymmetries in the velocity field lead to large residual
velocities with respect to the model velocity field. 
Additionally, they make it difficult to accurately determine the
inclination of the gas disk and we were forced to use the value from
the optical isophotal analysis. 
This all results in large uncertainties in the rotation curve,
especially in the outer parts.  
This galaxy is therefore not suitable for a derivation of the dark
matter properties and will not be used in our subsequent
mass-modelling. \\[0.2cm] 
{\bf UGC~2487} (NGC~1167) is a giant S0 galaxy ($M_B = -21.88,
D^B_{25} = 54 \, {\mathrm {kpc}}$; see \citetalias{Noordermeer06a}) with an
extended, highly regular gas disk.  
We can trace the \HI\ rotation curve out to radii of 80~kpc
(10~R-band disk scale lengths) and although there is a small decline
in the rotation velocities, they remain well above 300~km/s till the
outermost point.
The total mass enclosed within the last measured point is $M_{\mathrm
{enc}} = 2.1 \! \cdot \! 10^{12} \, \msun$, which makes UGC~2487 the
most massive galaxy in our sample.  
The total enclosed mass is larger even than those in the giant Sc
galaxies NGC~2916 and UGC~2885 \citep[][ note that in both papers a
Hubble constant of $H_0=50$~km~s$^{-1}$~Mpc$^{-1}$ is assumed; their
derived masses have to be divided by 1.5 when using our value of
75~km~s$^{-1}$~Mpc$^{-1}$]{Rubin79, Roelfsema85}; to our knowledge, it
is the largest mass ever derived from a rotation curve.   
\citet{Saglia88} list a number of other large disk galaxies with
extremely high rotation velocities; some of those galaxies may be even 
more massive than UGC~2487, but since no spatially resolved rotation
curves are available for these systems, no accurate values for the
total masses can be derived.   
In any case, UGC~2487 seems member of a class of extremely massive
disk galaxies \citep[see also][]{Giovanelli86,Carignan97}, with masses
that rival those of the most massive elliptical galaxies
\citep[e.g.][]{Bertin88,Minniti98}. \\
UGC~2487 is also classified as a Seyfert galaxy, explaining the broad
emission lines in the nucleus \citep[cf.][]{Filippenko85}. 
It has a central compact steep spectrum (CSS) radio source 
\citep[e.g.][]{Sanghera95,Giovannini01}, which is responsible for the 
\HI\ absorption in the centre.   
Away from the bright nucleus, we detect some very faint emission in
the optical spectrum. 
Although this emission seems to follow the general sense of rotation
of the galaxy, the emission profiles are broad and do not have
well-defined peaks.  
From these data alone, it is difficult to determine whether this faint
emission traces regular rotation in the circumnuclear regions, or
whether it is related to outflows from the active nucleus.  
Thus, this emission gives no useful information on the shape of the
potential in the inner regions and we have decided not to use it in
the derivation of the rotation curve. 
A small \HII region is detected 30\arcsec\ away from the
centre on the approaching side; the emission from this region has
regular line profiles and its velocity is consistent with the rotation
velocities of the \HI\ at the corresponding location. \\[0.2cm]
{\bf UGC~2916} has a regular, symmetric rotation curve in its central
regions. 
In the outer parts, however, a strong asymmetry is present between the
approaching and the receding sides of the galaxy. 
At the approaching side, the rotation velocities show a strong
increase outside the optical disk and reach a maximum of approximately
240~km/s at 20~kpc; further outwards the rotation velocities decline
again. 
At the receding side, the rotation velocities do not rise at all
outside the optical disk, but start to decline immediately.
It seems likely that the lopsidedness in this galaxy is the result of
the interaction with its nearby companion PGC~14370. \\[0.2cm]   
{\bf UGC~2953} (IC~356) is by far the best resolved galaxy in our
sample, with 117~independent data points in its rotation curve. 
The inner points, from the optical spectrum, sample the rotation
velocities at intervals of $\sim$80~pc or 0.02~R-band disk scale
lengths, whereas the last measured point lies at a projected radius of
59~kpc (14~disk scale lengths).   
The central rise in the rotation curve is unresolved even in the
optical spectrum; the rotation velocities rise to $\sim \! 200$~km/s
within 1 arcsecond from the centre, and keep rising more gradually from
there to a maximum of 310~km/s at $R \! \approx \! 100\arcsec$ ($\approx
7.5\, {\mathrm{kpc}}$, 2~disk scale lengths).  
In the outer regions, the rotation curve declines, with the asymptotic
velocity about 15\% lower than the maximum. 
The strong variation in the fitted inclination angles around $R = 600
\arcsec$ is probably an artefact caused by streaming motions in the
large spiral arm in the western parts of the galaxy; it was not judged
to be real and was therefore not used in the final tilted ring
fits. \\[0.2cm]
{\bf UGC~3205} has one of the most symmetric \HI\ disks of all
galaxies in our sample, both in its morphological appearance (see 
\citetalias{Noordermeer05}) and in its kinematics. 
The optical image is highly regular as well
(\citetalias{Noordermeer06a}). 
There is an almost perfect symmetry in the velocity field, and the
rotation curves for the approaching and receding sides of the galaxy
are identical within the measurement errors. 
The residual velocities are small too; the only significant residuals
are detected in the regions where the bar causes non-circular
streaming motions.   
It seems plausible that such streaming motions are also responsible
for the apparent variation in inclination angle in the central parts; we find 
it unlikely that the fitted variation is real and have assumed a
constant value for the final tilted ring fits. 
Outside the bar region, the residual velocities are of order 10~km/s
or less everywhere, indicating that the gas motion is highly regular
and undisturbed. \\
In the centre of the optical spectrum, a peculiar offset is detected between
the centre of symmetry of the rotation curve and the centre of the continuum
emission; the former is shifted by about one arcsecond ($\sim 250$~pc) to the
south-west with respect to the latter.
It seems unlikely that this offset is caused by absorption of light by
dust (as is most likely the case in UGC~3580), as the optical image is
highly symmetric and shows no signs of dust extinction whatsoever. 
No other peculiarities are seen in this galaxy at all and the origin
of the offset remains unclear to us. 
For the derivation of the optical rotation curve, we have used the
centre of symmetry of the line emission, rather than the continuum
centre, to fold the two halves onto each other. 
Given the flatness of the rotation curve and the high degree of
symmetry at larger radii in the spectrum, a different choice for the
dynamical centre would not have led to significantly different
rotation velocities, except for the very inner points. \\
The resulting rotation curve of this galaxy seems to lack, almost 
completely, the characteristic steep rise in the centre which is
observed in most other galaxies in our sample. 
Parts of the central regions are devoid of gas though, and we cannot
trace the entire rise of the rotation curve. \\[0.2cm]   
{\bf UGC~3546} (NGC~2273) is classified as a Seyfert galaxy
\citep{Huchra82}.  
High-resolution observations with the VLA revealed a bright radio
continuum source in the centre, consisting of two separate lobes
separated by about 0.9\arcsec\ or $\approx \! 120$~pc
\citep{Ulvestad84,Nagar99}.   
The optical spectrum shows strong nuclear emission lines, the nature of
which has been discussed by \citet{Ho97}. 
The \Ha emission in our optical spectrum is clearly extended
\citep[cf.][]{Pogge89a,Mulchaey96}, but the line profiles are
irregular and broad out to a radius of about 10\arcsec; within this radius, it
is difficult to disentangle the effects of quiescent rotation from possible
in- and outflows from the nucleus.  
The uncertainties in the inner points of the rotation curve are
therefore large, and the sharp peak and subsequent decline in the
rotation curve may not be real. \\  
In contrast, the \HI\ velocity field displays smooth and regular
rotation throughout the entire gas disk of the galaxy. 
The central part of the velocity field appears somewhat distorted, but this is
an artefact caused by beam smearing \citepalias[see also][]{Noordermeer05}.   
This also explains the erratic behaviour of the fitted orientation
angles in the inner regions; for the final tilted ring fits, we have
used the position angle and inclination determined from larger radii,
as indicated with the bold lines in the figures in appendix~\ref{app:figures}.
No effects can be seen of non-circular motion in the bar, possibly due
to its favourable orientation (perpendicular to the major
axis). \\[0.2cm]  
The centre of symmetry in the optical spectrum of {\bf UGC~3580} has
an offset of approximately one arcsecond ($\sim \! 100$~pc) with
respect to the peak in the continuum emission (indicated with the
dashed line in the figure in appendix~\ref{app:figures}). 
This difference is most likely caused by obscuration of the continuum
emission by dust. 
As was noted in \citetalias{Noordermeer06a}, the optical image shows
strong dust features in the central regions of this galaxy. 
Thus, a determination of the centre of the galaxy based on the peak in
the light distribution is problematic and it is not surprising that
the dynamical centre is offset with respect to the isophotal one. \\ 
Not only the optical appearance of this galaxy is peculiar, the
kinematical structure is remarkably different from that of most
galaxies in our sample as well. 
There is a marked asymmetry between the approaching and receding
sides of the optical spectrum. 
Whether this reflects a true lopsided kinematics in the central regions or is
a result of dust extinction too, is difficult to determine from these
data alone. 
The major axis slice through the \HI\ data cube is also asymmetric,
but it appears that this asymmetry is peculiar to the major axis; it
does not occur equally strong at different position angles and the
rotation curves averaged over the full halves of the velocity field
are only marginally different.\\ 
Similarly to other galaxies in our sample, the central rise of the
rotation curve of UGC~3580 is steep. 
But where in most of our galaxies the rotation curve becomes flat
after the initial rise, or even starts to decline, it continues to
rise gradually till twice the optical diameter ($\sim \! 8.5$~R-band
disk scale lengths) in this galaxy and only flattens out at the edge of the
gas disk.
This behaviour resembles that of typical late-type and dwarf galaxies
\citep{Broeils92,Swatersthesis} which generally have slowly rising
rotation curves too.
However, although  UGC~3580 is one of the least luminous galaxies
in our sample ($M_{\mathrm B} = -18.31$) and correspondingly has one
of the lowest maximum rotation velocities ($V_{\mathrm {max}} = 127 \,
{\mathrm {km/s}}$), its luminosity is still too high to classify it as
a dwarf galaxy.    
UGC~3580 thus seems to be a relatively luminous member of a
class of low-luminosity early-type disk galaxies which have distinctly
different morphological and kinematical features compared with their
high-luminosity counterparts; other nice examples of this class are
UGC~6742, UGC~12043 and UGC~12713 (see \citetalias{Noordermeer05}).
\\[0.2cm]    
{\bf UGC~3993} is an S0 galaxy with a regularly rotating gas disk. 
It resembles UGC~2487, but it is not as large nor as massive. 
Although the galaxy is quite face-on, the \HI\ velocity field is of
sufficient quality to determine the inclination with reasonable
accuracy. 
Nevertheless, the uncertainties in the rotation velocities due to the
inclination errors are large. 
In particular, we cannot exclude the possibility that the decline in
the rotation curve of this galaxy is caused by a small warp in the
outer parts towards a more face-on orientation. \\[0.2cm]
The rotation curve of {\bf UGC~4458} (NGC~2599) looks remarkably
different from the canonical flat rotation curves normally observed in
spiral galaxies. 
Instead, it rises very rapidly to an extreme peak velocity of 490
km/s, after which it shows a sharp decline of more than 50\% before
it asymptotically approaches a constant rotation velocity of $\sim \!
240$~km/s.   
The central rise in the rotation curve is unresolved even in our
optical spectrum; it is possible that in the very inner regions, gas
moves at even higher rotation velocities. 
Already, a rotation velocity of 490 km/s is unusually high, and seems
surpassed only by that of UGC~12591 \citep{Giovanelli86}. \\
UGC~4458 is, however, close to face-on and the errors in the
rotation velocities, caused by the uncertainties in the exact value of 
the inclination, are large. 
Thus, the peak velocity could be substantially lower if the galaxy
were slightly more inclined in the inner parts, and the extreme
decline in the rotation curve could partly be explained by a small
warp towards a more face-on orientation in the outer parts. 
To fully explain the decline as a result of a warped gas disk, however, the 
inclination would have to decrease steadily from 31\deg\ in the centre to
15\deg\ in the outer parts (assuming a constant rotation velocity of
400~km/s).   
Such a change in inclination angle is not detected in the
tilted ring fits, and it seems unlikely that the decline in the
rotation curve can be explained fully by warping of the gas
disk. 
Meanwhile, we cannot exclude a small warp towards a higher inclination in the
outer regions either. 
In particular, our data are also consistent with a continuing, Keplerian
decline in the outer points of the rotation curve, and thus, with an absence
of dark matter inside the radii probed by the \HI\ disk. 
We will investigate this issue in more detail in our subsequent paper on the
mass modelling.  \\[0.2cm]  
The central parts of the gas disk of {\bf UGC~4605} (NGC~2654) are
close to edge-on, with an estimated inclination angle of 84\deg.    
At larger radii, the gas disk warps towards a more face-on
orientation.   
Due to the high inclination and the resulting line-of-sight
integration effects, the central regions of the velocity field have a
bias towards the systemic velocity and do not give an accurate
representation of the projected rotational velocities (see also
\citetalias{Noordermeer05}), so the usual tilted ring analysis could not be
used there.     
Within a radius of $R = 30\arcsec$, this problem could be circumvented
by using the optical spectrum; due to its higher spatial resolution,
it suffers much less from projection effects. 
Between $R = 30\arcsec$ and $R = 110\arcsec$, where no optical
information is available, we determined the rotation velocities by
hand from the \HI\ data, using the same method that was normally used 
for the inner regions of the optical spectra
(section~\ref{subsec:Harotcurs}).  
At each position along the major axis, the terminal velocity of the
line profile was determined and assumed to represent the projected
rotational velocity at the line of nodes; the average of the rotation
velocities for the approaching and receding sides of the galaxy was
then taken as the true rotation velocity at that radius. 
Outside $R = 110\arcsec$, the gas disk becomes sufficiently less
inclined that tilted ring fits could be applied to the velocity
field; the rotation curve for $R > 110\arcsec$ was therefore
determined in the usual way. \\
The edge-on orientation of this galaxy also has a positive effect, 
namely that the uncertainties in the inclination angle are small; the
resulting errors in the rotation velocities are almost negligible. \\ 
The final rotation curve is well defined and symmetric, except for the
region around 120\arcsec, where the rotation velocities on the
approaching side are declining already while those on the receding
side are still constant. 
Beyond this region, the rotation curve is symmetric and shows a strong
decline on both sides. 
At larger radii ($R \ga 200\arcsec$), the rotation curve flattens out
at a level of approximately 185~km/s. \\
Due to the edge-on orientation of this galaxy, no accurate optical
photometry could be obtained and we were only able to obtain an
estimate for the radial scale length of the stellar disk. 
But the rotation curve is relatively flat between 2 and 3 times the 
estimated scale length, so the value for $V_{\mathrm {2.2h}}$ is
reasonably robust. 
Due to the lack of detailed information on the stellar mass
distribution, however, this galaxy can not be used for mass
modelling. \\[0.2cm] 
The outer parts of the gas disk of {\bf UGC~5253} (NGC~2985) are
dominated by a large spiral arm extending from the north of the
galaxy.  
Although the gas in the arm is clearly rotating, it is impossible to
determine the exact orientation of the arm and we have only fitted a
tilted ring model to the inner parts. 
Even so, our rotation curve extends out to a radius of 49~kpc, or
$\sim \! 9$ R-band disk scale lengths. \\
The most interesting aspect of UGC~5253, however, is the strong m=0
component in the residual velocity field. 
Both in the inner ($ R \la 200\arcsec$) and in the outer ($ R \ga
400\arcsec$) regions, the residual velocities are small, indicating
that our fitted tilted ring model is an accurate description of the
observed gas motions.   
Around a radius of 300\arcsec ($\approx 30$kpc), however, a ring-like feature
is detected in the residual field with an amplitude of $-20$~km/s.
This feature has a high degree of symmetry with respect to the centre
of the galaxy and the residual velocities are almost constant with
position angle; only at the western end of the ring are the residuals
slightly lower ($\sim \! -10$~km/s). 
In the rotation curve, the feature manifests itself as a marked
asymmetry between the approaching and receding side. 
However, simple kinematical asymmetries can only explain residuals
along the major axis of a galaxy; without additional radial motions,
they must always vanish on the minor axis.    
The fact that the residual velocities in the ring are so symmetric and also
present on the minor axis, argues against a simple explanation in terms of a
kinematical asymmetry. \\ 
A more plausible explanation of the marked m=0 component in the
residual field is that it is induced by the gravitational
perturbations from the large spiral arm in the outer parts. 
It was shown by \citet{Schoenmakers97} on theoretical grounds that an
m=1 perturbation in the gravitational potential leads to a strong m=0
term in the residual velocity field. 
The combination of the large m=1 spiral arm in the gas distribution of
UGC~5253 and the pronounced m=0 term in its residual field are thus a
strong empirical confirmation of their predictions. 
Furthermore, \citet{Schoenmakers97} showed that it is, in principle,
possible to use the amplitude of the m=0 term in the residual field to
measure the strength of the perturbation on the potential. 
This analysis is, however, beyond the scope of this paper and will be
postponed to a later time. \\
Alternatively, we could be seeing a vertical vibrational mode in the
gas disk of UGC~5253, where the entire ring is moving upwards (i.e.\
towards us) with respect to the rest of the galaxy. 
\citet{Sellwood96} and \citet{Edelsohn97} showed the results of N-body 
simulations which suggest that vertical vibrations can exist in the
disks of galaxies, possibly triggered by a tidal interaction with a
companion galaxy. 
It seems, however, questionable if such a mechanism could explain the
highly symmetric feature we observe in UGC~5253, and we judge the
explanation of \citet{Schoenmakers97} more plausible. \\
Note that the feature in the residual field coincides with a marked
drop in the rotation curve.
Both inside $R = 200\arcsec$ and outside $R = 400\arcsec$, the
rotation curve is flat, but around $R = 300\arcsec$, it suddenly drops
from 245 to 210~km/s; the rate of the decline is consistent with pure
Keplerian decay. 
Whether it is a coincidence that this drop occurs at similar radii as
the feature in the residual field, or whether both effects are
related, is unclear.  \\[0.2cm]   
A strong stellar \Ha absorption feature is present in the central
arcseconds of the optical spectrum of {\bf UGC~6786} (NGC~3900); no
\Ha emission is detected in the inner parts. 
Emission {\em is} detected in the central parts in the 6583.46~\AA\ 
[N{\sc ii}] line, but the standard procedure of stacking together the
different emission lines in the optical spectrum cannot be used here.
Instead, we analysed the \Ha and [N{\sc ii}] lines separately in this
case, and combined the resulting rotation curves afterwards; at
positions where emission was detected in both lines, the average
velocity was calculated. 
The spectrum shown in the figure in appendix~\ref{app:figures}
was created by replacing the inner part (15 pixels on either side of
the centre) of the \Ha spectrum by the corresponding region of the
[N{\sc ii}] line.  
Thus, it shows the extended \Ha emission in the outer regions,
together with the [N{\sc ii}] emission in the nuclear region. \\
Note that the central emission in the [N{\sc ii}] line is irregular,
with broad line profiles. 
This may be explained as a result of line-of-sight integration effects
through the inner regions of the massive bulge of UGC~6786 (see
\citetalias{Noordermeer06a}), but observations at higher spatial
resolution and sensitivity are required to investigate this in more
detail. 
At a radius of 5\arcsec\ on the approaching side, a strong emission
feature is detected which has a velocity that lies more than 100~km/s
closer to the systemic velocity than the emission at smaller radii. 
It seems unphysical that this emission traces regular rotation of gas
in the plane of the galaxy and we did not include it in the combined
rotation curve. \\
The outer parts of the \HI\ disk are distorted as well, with two large
spiral arms extending from the symmetric inner disk. 
It is impossible to determine the exact orientation of the gas in
these arms and no tilted rings were fit outside a radius of
240\arcsec. 
The residual velocity field shows a peculiar m=5 harmonic component in
the azimuthal direction. 
According to the results of \citet{Schoenmakers97}, this implies an m=6
perturbation in the gravitational potential of this galaxy. 
No obvious source of such a perturbation can be identified in the visible
matter in this galaxy, so the perturbation, if real, must be caused by the
dark matter halo. \\ 
Since UGC~6786 does not have a regular exponential stellar disk, no
disk scale lengths are indicated in the figures in
appendix~\ref{app:figures}. \\[0.2cm]  
The rotation curve of {\bf UGC~6787} (NGC~3898) is well resolved and 
shows some characteristic `wiggles' with an amplitude of 30 --
50~km/s.   
The kinematics in the central parts are only barely resolved in the
optical spectrum, and due to the high inclination angle and resulting
line-of-sight integration effects, the central line-profiles are
strongly broadened. 
After the initial rise of the rotation curve to the peak velocity of
270~km/s, the rotation velocities drop to approximately 220~km/s at a
radius of 30\arcsec\ ($\sim \! 2.75$~kpc), after which they gradually rise  
again to 250~km/s at $R\approx100\arcsec$ ($\sim \! 9$~kpc). 
The rotation velocities then drop again to 220~km/s, after which they
rise again to reach a more or less flat plateau at 250~km/s. 
Although there are clear indications that the gas disk of UGC~6787 is
warped, the locations of the `wiggles' in the rotation curve do not
coincide with the radii where the position angle and the inclination
change and the variations in the rotation velocity seem real. 
This is further confirmed by the fact that the variations occur
symmetrically at all position angles over the velocity
field. \\
The discrepancy between the kinematic inclination angle derived here and the
optical inclination from \citetalias{Noordermeer06a} can be explained by the
dominance of the bulge in the optical image.  
As was noted in \citetalias{Noordermeer06a}, the optical image of this galaxy
is dominated by the spheroidal bulge out to large radii, which makes it
impossible to obtain a reliable estimate for the inclination from the
isophotes.
The kinematical inclination derived here is free of such effects and
thus reflects the true orientation of this galaxy more
accurately. \\[0.2cm]    
{\bf UGC~8699} (NGC~5289) is highly inclined and the central line
profiles in the optical spectrum are severely broadened by the
combined effect of beam smearing and line-of-sight integration
effects.  
Similar to UGC~6787, the rotation curve of this galaxy shows a
distinct `bump'. 
After an initial rise, the rotation curve reaches a peak velocity of
205~km/s at a radius of approximately 8\arcsec ($\approx \! 1.4$~kpc).  
At $R\approx 30\arcsec$ ($\approx \! 5.5$~kpc), the rotation
velocities have fallen to 170~km/s, but unfortunately, no \Ha emission
is detected between 15 and 30\arcsec\ on either side of the galaxy, so
the exact shape of the decline in the rotation curve cannot be
recovered.   
Between $R=30\arcsec$ and $R=45\arcsec$, the rotation curve rises
back to approximately 200~km/s, symmetrically on both sides of the
galaxy. 
At the approaching side of the optical spectrum, no emission is
detected anymore beyond this radius; at the receding side, the
spectrum indicates a small decline again in the rotation velocities,
out to the last measured point at 54\arcsec ($\approx \! 9.7$~kpc). \\ 
In the \HI\ observations, UGC~8699 is poorly resolved along the minor axis.  
Comparison of the xv-slice through the \HI\ data cube with the optical
spectrum shows that there is probably a central hole in the \HI\ disk,
explaining the absence of \HI\ emission at high rotation velocities
close to the centre.  
Thus, this galaxy provides another good illustration of the use of the 
optical observations to resolve the shape of the inner rotation curve.  
Because of the poor resolution along the minor axis, standard tilted
ring fits did not recover the rotation velocities accurately in this
case. 
Instead, we used only points close to the major axis to determine the
\HI\ rotation curve; points within 60\deg\ of the minor axis were
discarded in the fits. 
In the outer parts, a small asymmetry is detected between the rotation
velocities of the approaching and receding sides of the galaxy, with
the former rotating about 10\% slower than the latter. \\[0.2cm] 
{\bf UGC~9133} (NGC~5533) has the most extended rotation curve in our
sample, with the outermost point in the rotation curve lying at a
projected radius of 103~kpc ($>$ 11 R-band disk scale lengths).  
With a rotation velocity at this radius of 225~km/s, the total
enclosed mass is $1.3\cdot10^{12} \, \msun$. 
Most gas at large radii lies in a giant spiral arm which extends from
the north-east side of the gas disk and is warped with respect to the
inner parts of the galaxy.  
Although the arm is clearly rotating and the fitted position and
inclination angles are well-behaved, the lack of symmetry in the arm
makes it difficult to exclude the possibility that the gas in the arm
is not rotating at perfectly circular orbits.
Therefore, care should be taken with the rotation velocities outside
R $\approx 200\arcsec$.   
Note, however, that we have assumed rather conservative values for the
uncertainty $\Delta i$ in the inclinations, such that the
corresponding uncertainties $\Delta V_i$ in the rotation curve
(indicated with the shaded area in the figures in
appendix~\ref{app:figures}) also include, at least partly, the
uncertainties introduced by the orientation of the spiral arm.
Note also that the residual velocities with respect to the tilted ring
model are small, indicating that the non-circular motions are not
dominant. \\
The optical observations do not have sufficient resolution to trace
the rise of the rotation velocities in the centre. 
Furthermore, a bright central component complicates the interpretation
of the spectrum in the inner few arcseconds (see also
appendix~\ref{app:rotcurs_centres}).  
At larger radii, however, the spectrum is highly regular and
symmetric.  \\
The rotation curve of UGC~9133 keeps declining almost to the
outermost points in the \HI\ rotation curve. 
The asymptotic velocity is about 25\% lower than the maximum. 
Only at a radius of approximately 80~kpc does the rotation curve
flatten out. \\[0.2cm]
{\bf UGC~11670} (NGC~7013) has a giant bar; the large `wiggle' in 
the rotation curve around R = 50\arcsec\ may well be an artefact
caused by streaming motions in this bar.  
The change in position angle which is detected at those radii is most
likely also caused by non-circular motions and for the final rotation
curve, we assumed that the position angle is constant at -24\deg\ in
the inner regions. 
At larger radii, the rotation curve becomes smoother and shows a
gradual decline out to a radius of about 300\arcsec (18~kpc), after
which the rotation velocities seem to rise again. 
The upturn is only visible in the low-resolution data, but seems to
occur on both sides of the galaxy, which strengthens the detection. 
We can, however, not exclude the possibility of in- and outflows in
the outer parts of the gas disk of this galaxy; deeper observations
are needed to verify if the gas in the outer parts is truly on
circular orbits and if the rise in the rotation curve is real. \\
In the centre, a bright nuclear component complicates the
interpretation of the optical spectrum (see
appendix~\ref{app:rotcurs_centres}) and the inner points of the  
rotation curve may not be reliable. 
\\[0.2cm] 
{\bf UGC~11852} has the most extended gas disk, relative to the
optical size, of all galaxies in our sample. 
It allows us to trace the rotation curve to a radius of $\sim \!
90$~kpc, or $\sim \! 21$~R-band disk scale lengths. 
This galaxy is clearly warped: the fitted position angles from the
tilted ring fits change by more than 20\deg\ from the inner to the
outer parts of the disk.
The inclination changes as well and the galaxy becomes more edge-on in
the outer parts.  
The apparent variation in the fitted inclinations in the inner regions
is probably an artefact caused by non-circular motions in and around
the bar; it is not included in the final fit.   
The rotation curve can be seen to rise steeply in the inner regions, although 
the exact shape of the inner rotation curve cannot be recovered due to the
strong central component present in the optical spectrum (see also
appendix~\ref{app:rotcurs_centres}).  
Outside the optical disk, the rotation curve shows a marked decline,
but it seems to flatten out in the outer parts. 
The asymptotic rotation velocity is approximately 25\% lower than the
maximum. 
Strong residual velocities (up to $-50$~km/s) are detected
northwest of the centre of the galaxy.  
The velocities of the gas in this region are incompatible with regular
rotation and indicate large-scale streaming motions. \\[0.2cm]
{\bf UGC~11914} (NGC~7217) has a highly regular gas disk; the rotation
curves of the approaching and receding sides separately are almost
indistinguishable.   
Unfortunately, however, the inclination is not tightly constrained and
the resulting uncertainties in the rotation curve are large. \\
Comparison of this galaxy with similar systems in our sample (e.g.\
figures~\ref{fig:casertanogram} -- \ref{fig:binnedrotcurs}) implies that it
should have a declining rotation curve. 
The fact that we do not see such a decline is probably caused by the small
radial extent of the \HI\ disk, which enabled us to measure the rotation curve
out to 3.3 R-band disk scale lengths only.   
In several other galaxies in our sample (e.g.\ UGC~2487, 5253), the
decline in rotation velocities sets in around the edge of the optical
disks, and it does not seem unreasonable to assume that the rotation curve
of UGC~11914 declines at similar radii. \\
Due to its relative proximity, UGC~11914 is one of the few galaxies
where we fully resolve the inner rise of the rotation curve in the
optical spectrum. 
We clearly see the gradual change in velocity from the approaching to
the receding side over the central 30\arcsec. 
Superposed over the `normal' emission, we detect the bright and unresolved
central LINER component \citep[cf.][]{Keel83}.
The line profile of the LINER is very broad and implies velocities well in 
excess of the quiescent rotation velocities at larger radii (see also
appendix~\ref{app:rotcurs_centres}). \\[0.2cm]   
{\bf UGC~12043} (NGC~7286) is the least luminous galaxy in our
rotation curve sample and the amplitude of the rotation curve is
correspondingly low ($V_{\mathrm {max}} = 95 \, {\mathrm {km/s}}$).  
In contrast, the rotation curve is the second most extended, compared 
to the optical size, of all galaxies in our sample, with the outermost
point lying at a projected radius almost 19 times larger than the
R-band disk scale length.  
This galaxy is the only case in our sample with a slowly rising
rotation curve.
It completely lacks the steep central rise observed in all other
galaxies in our sample and, instead, rises gradually to the maximum
which is only reached around 3 optical scale lengths.  
This behaviour can be explained by the fact that UGC~12043 has no
bulge and thus has a much smaller central surface density than all
other galaxies in our sample (see \citetalias{Noordermeer06a}). 
UGC~12043 is thus a member of a class of low-luminosity early-type disk 
galaxies which have distinctly different morphological and kinematical
features than their high-luminosity counterparts; other nice examples
of this class are UGC~3580, UGC~6742, and UGC~12713
\citepalias[see][]{Noordermeer05}. \\
The interpretation of the optical spectrum is complicated by the low 
spectral resolution of the GoldCam spectrograph used for these
observations.\label{page:ugc12043}
Although the rotation of the galaxy can clearly be detected, the line
profiles are practically unresolved and no corrections can be made
for beam smearing or line-of-sight integration effects.  
A sudden jump occurs in the receding side of the optical spectrum,
around R=20\arcsec, the nature of which is unclear; no emission is
detected at the corresponding radius on the approaching side.\\ 
The \HI\ velocity field is highly symmetric, and the inferred rotation
curves for the approaching and receding sides separately are
virtually indistinguishable, except for the outer parts of the
low-resolution velocity field (not shown in the figure in
appendix~\ref{app:figures}), where the approaching side
exhibits a sudden rise in the rotation velocity. This rise is
probably caused by non-circular motions in the outer disk of the
galaxy and most likely does not correspond to a real increase in the
rotation velocity; it is not included in the final, azimuthally
averaged rotation curve.

\section{Broad central velocity profiles in the optical spectra}
\label{app:rotcurs_centres}
In about half of the galaxies in our sample, broad velocity profiles
are present in the optical spectra of the central regions; the 
velocity amplitudes of those features often exceed the rotational
velocities observed at larger radii.
In some cases, especially in the Seyferts UGC~2487 and 3546, these
broad components may be explained as a result of nuclear activity and
related gas flows. 
In the other cases, however, no strong nuclear activity is observed
and the most natural explanation of the observed velocities is a
nuclear disk or ring, possibly rotating around a massive nuclear
star cluster or super-massive black hole. 
Similar rapidly rotating components have been observed in the centers
of many other spiral galaxies (e.g.\ \citealt{Carter93, Bertola98,
Sofue98, Sofue99, Sofue01} and references therein;
\citealt{McDermid04, Fathi04}), and may well be the spectroscopic 
counterparts to the nuclear structures imaged by e.g.\
\citet{Carollo97, Carollo98}.  
\begin{table*}
 \begin{minipage}{12.cm}
  \centering
  \caption[Central features in the optical spectra]
  {Central features in the optical spectra. (1)~UGC~number;
   (2)~central velocity of line profile; (3)~profile width, measured at 
   20\% of the maximum flux; (4)~derived rotational velocity;
   (5)~estimated diameter; (6)~total enclosed mass; (7)~average density
   within $D_c$; (8)~total R-band luminosity within $D_c$ and
   (9)~R-band mass-to-light ratio within $D_c$.
   \label{table:centralstructures}}  
      
  \begin{tabular}{r|rccd{1.1}d{1.2}d{1.2}d{1.3}d{1.2}c}
   \hline 
   \multicolumn{1}{c|}{UGC} & \multicolumn{1}{c}{$V_0$} &
   $W_c$ & $V_c$ & \multicolumn{2}{c}{$D_c$} &
   \multicolumn{1}{c}{$M_c$} & \multicolumn{1}{c}{$\rho_c$} &
   \multicolumn{1}{c}{$L_{c,R}$} & $(M/L)_{c,R}$ \\  
    
    & \multicolumn{1}{c}{km/s} & km/s & km/s & 
   \multicolumn{1}{c}{\arcsec} & \multicolumn{1}{c}{kpc} &
   \multicolumn{1}{c}{$10^9 \, \msun$} & \multicolumn{1}{c}{$10^3
   \frac{\msun}{\mathrm{pc}^{3}}$} &  
   \multicolumn{1}{c}{$10^8 \, \lsun$} & \msun/\lsun \\ 

   \multicolumn{1}{c|}{(1)} & \multicolumn{1}{c}{(2)} &
   (3) & (4) & \multicolumn{2}{c}{(5)} & \multicolumn{1}{c}{(6)} & 
   \multicolumn{1}{c}{(7)} & \multicolumn{1}{c}{(8)} & (9) \\    
   \hline 
   
    2487 & \multicolumn{5}{l}{\hspace{0.5cm} -- Seyfert nucleus --} \\
    2953 &  911 & 470 & 307 & 1.1 & 0.08 & 0.86 & 3.4   & 0.34 & 25 \\
    3546 & \multicolumn{5}{l}{\hspace{0.5cm} -- Seyfert nucleus --} \\
    5253 & 1324 & 317 & 263 & 1.1 & 0.11 & 0.88 & 1.3   & 1.3  &  7 \\
    8699 & 2553 & 434 & 227 & 1.1 & 0.20 & 1.2  & 0.28  & 0.58 & 21 \\
    9133 & 3860 & 490 & 307 & 1.5 & 0.40 & 4.4  & 0.13  & 5.7  &  8 \\
   11670 &  781 & 357 & 190 & 2.6 & 0.16 & 0.69 & 0.30  & 0.96 &  7 \\
   11852 & 5851 & 355 & 232 & 1.9 & 0.74 & 4.6  & 0.022 & 4.5  & 10 \\
   11914 &  947 & 472 & 458 & 1.1 & 0.08 & 2.0  & 6.9   & 0.42 & 48 \\
   \hline
  \end{tabular}
 \end{minipage}
\end{table*}  

In most cases, the central structures are spatially unresolved in our
spectra, but in a few cases (e.g.\ UGC~11670, 11852) the resolution is
sufficient to detect a velocity gradient. 
These gradients are always in the same  direction as the sense of
rotation of the gas at larger radii, which further strengthens our
assumption that the central features in the spectra originate from
regularly rotating gas. 

The data presented here lack the spatial resolution to unambiguously
determine the nature of the central components in our spectra. 
A proper investigation would require sub-arcsecond resolution, both
for the kinematic as for the photometric data, i.e.\ either
space-based or adaptive-optics assisted observations. 
A nice example of the potential of such observations was recently
presented by \citet{Atkinson05}, who used imaging and long-slit
spectroscopy from the Hubble Space Telescope to derive limits on the
mass of the central black holes in NGC~1300 and 2748 \citep[see
also][]{Harms94, Ferrarese99}. 

With our data, we can only estimate the total mass in the central
regions of our galaxies. 
The results of our crude analysis are summarized in
table~\ref{table:centralstructures}.  
The rotational velocities of the gas were derived from the width $W_c$
of the central line profiles, measured at 20\% of the maximum flux, 
assuming that the central gas has the same orientation as the outer
disk: $V_c = W_c / (2 \sin i)$. 
Note that the true rotational velocities may be larger or smaller if
the central disks are tilted with respect to the main disk for which
we derived the inclination angle, or if significant non-circular
motions are present.  
The total extent of the gas in the central component was estimated by
eye from the spectra directly;
table~\ref{table:centralstructures} gives the estimated 
diameters $D_c$. 
In most cases, the resolution of our data allows us to probe diameters
of a few hundred parsecs or less; for the distant galaxy UGC~11852,
the constraints are somewhat worse. 
Since in all cases, the structures are not or only marginally
resolved, the diameters given in
table~\ref{table:centralstructures} are highly uncertain and, strictly
speaking, upper limits only.  
Finally, if we make the additional assumption that the central mass
concentration has a spherically symmetric shape, the total enclosed
mass can be estimated using equation~\ref{eq:enclosedmass}.   

Table~\ref{table:centralstructures} clearly shows that the
inferred central masses and densities are high, especially in the
nearby galaxies where we have better constraints on the total extent
of the rapidly rotating gas. 
Central densities of order $10^3 \, \msun \, \mathrm{pc}^{-3}$ and
higher can almost certainly not be explained by the normal stellar
components. 
In column (8) of table~\ref{table:centralstructures}, we give
the total R-band luminosities within $D_c$, as measured from the
optical images presented in \citetalias{Noordermeer06a}. 
It is clear that the observed luminosities are too small to account
for the derived dynamical masses; the local mass-to-light ratios are
much larger than what is expected for the surrounding bulge material.   
 
Our inferred masses are also a few orders of magnitudes larger than
those of the most massive stellar clusters known to date
\citep[e.g.][]{Mengel02,Maraston04, Walcher05}, so it is unlikely that
the high rotation velocities are caused by unresolved central
concentrations of stars. 
For UGC~2953, 5253, 11670 and 11914, HST images are available
\citep[e.g.][]{Carollo02, Hunt04b} and we 
could explicitly verify that no bright and compact sources of light
are hidden in our own, lower-resolution images.

In conclusion, the central components in our spectra seem to be a
strong indication for the presence of super-massive black holes in at
least a fraction of our galaxies.   
Sub-arcsecond observations, however, are required to obtain more
detailed knowledge on the spatial extent and orientation of the
rapidly rotating gas and to provide conclusive evidence for the
presence and mass of the black holes.

\section{Rotation curves}
\label{app:figures}
On the following pages, we present the rotation curves and several
other related quantities for all galaxies in the sample. For each
galaxy, we show a figure consisting of the following panels: \\[0.25cm]
{\bf Left hand column:} velocity fields \\
{\em Top\/}: grayscale and contour representation of the observed \HI\
velocity field from \citetalias{Noordermeer05}. Darker shading and white
contours indicate the receding side of the galaxy. Contours are spaced at
intervals of 25 km/s; the thick contour indicates the systemic velocity from  
table~\ref{table:data}. \\ 
{\em Middle\/}: model velocity field, based on the tilted ring fits to the
velocity field shown above. Contours and grayscales are the same as in the
observed velocity field. \\  
{\em Bottom\/}: Residual velocity field, produced by subtracting the model
velocity field from the observed one. The grayscales range from -40 to +40
km/s (white to black). Contours are spaced at intervals of 25 km/s; the 
thick contour indicates zero residual velocity. \\ 
The cross in each panel indicates the dynamical centre, as given in
table~\ref{table:data}.  \\[0.25cm] 
{\bf Top row:} x-v diagrams \\
{\em Middle\/}: cleaned and stacked optical spectrum (see
section~\ref{subsubsec:optredux} for a description of how these were
produced). \\  
{\em Right\/}: major axis position-velocity slice through the \HI\ data
cube. \\  
The position angles on the sky of both plots are indicated in the top left
corner of the panels. Contours in both panels are at -1.5 and -3 (dotted) and 
$1.5,3,6,12,\ldots$ times the rms noise in the respective datasets. The dashed
horizontal and vertical lines denote the systemic velocity and the centre of
the galaxy respectively. The fitted rotation velocities are overplotted. For
each of the two sides of the plots, approaching and receding, we plot the
velocities derived from fits to that side only. Plotting symbols are as
follows:  
\begin{list}{--}{\leftmargin=0.cm \itemsep=0.cm \parsep=0.cm
    \topsep=-0.2cm} 
 \item blue/black squares show velocities derived from the final tilted ring
   fits to the \HI\ velocity fields; 
 \item red/dark gray bullets show velocities fitted to the optical line
   profiles.  
 \item orange/light gray bullets show the central locations in the optical
   spectra which were affected by `optical beam-smearing' and where the
   rotational velocities were determined by eye, rather than by Gaussian fits
   (see section~\ref{subsubsec:optbeamsmear}).  
\end{list} 
\vspace{0.2cm}
{\bf Bottom row, middle panels:} Orientation parameters from the tilted ring
fits to the \HI\ velocity fields. \\
{\em Top\/}: data points show the fitted position angles (north through east)
from fits with position and inclination angle left free.  
The bold lines give the values that were used for the final fits to
derive the rotation curves. 
The arrows give the values derived in \citetalias{Noordermeer06a} from the
outer isophotes of the optical image. \\  
{\em Bottom\/}: data points show the fitted inclination angles from fits with
inclination angle left free and position angle fixed at the values shown with
the bold line in the top panels.  
The bold lines give the values for the inclination angle that were
used for the final fits to derive the rotation curves; the shaded
regions show the adopted uncertainties.  
The arrows give the values derived from the outer isophotes of the
optical image.\\[0.25cm]
{\bf Bottom row, right hand panels:} Rotation curves. \\
{\em Left\/}: inner regions. \\
{\em Right\/}: full rotation curves. \\
The axes at the bottom show radii in arcseconds, those at the top show the
corresponding radii in kiloparsecs. \\ 
Plotting symbols and linestyles are as follows:
\begin{list}{--}{\leftmargin=0.cm \itemsep=0.cm \parsep=0.cm
    \topsep=-0.2cm} 
 \item filled blue/black squares show velocities derived from the final tilted
   ring fits to the \HI\ velocity fields; the position and inclination angles
   assumed for these final fits are indicated with the bold lines in the middle
   panels of the bottom row.  
 \item open blue/black squares show velocities that were derived from \HI\ data
   at lower resolution than those shown in the left column and top row. 
 \item red/dark gray bullets show the velocities from the optical spectra.
 \item orange/light gray bullets show optical velocities that were manually
   adapted to correct for beam-smearing and line-of-sight integration effects
   in the optical spectra (see section~\ref{subsubsec:optbeamsmear}).
 \item crosses and plus-signs indicate the rotation curves for the approaching
   and receding sides respectively.  
 \item errorbars are a combination of fitting errors and differences between
   the approaching and receding sides (see section~\ref{subsec:errors}).  
 \item bold lines show the smoothed rotation curves, derived from cubic spline
   fits through the individual data points (see
   section~\ref{subsec:finalsteps} for details).   
 \item dashed lines indicate regions where no optical emission was detected. 
 \item shaded regions show the uncertainties in the rotational velocities due
   to the adopted uncertainties in the inclination; for clarity, they are
   drawn around the smoothed rotation curve, rather than around the individual
   data points.
 \item vertical arrows at the bottom show the radii corresponding to one
   (left) and two (right) disk scale lengths of the R-band image
   \citepalias[see][]{Noordermeer06a}; horizontal arrows at the right show the
   derived maximum and asymptotic rotation velocities and the rotation
   velocity at 2.2 R-band disk scale lengths (see table~\ref{table:data}).
\end{list}

\clearpage

\begin{figure*}
    {\psfig{figure=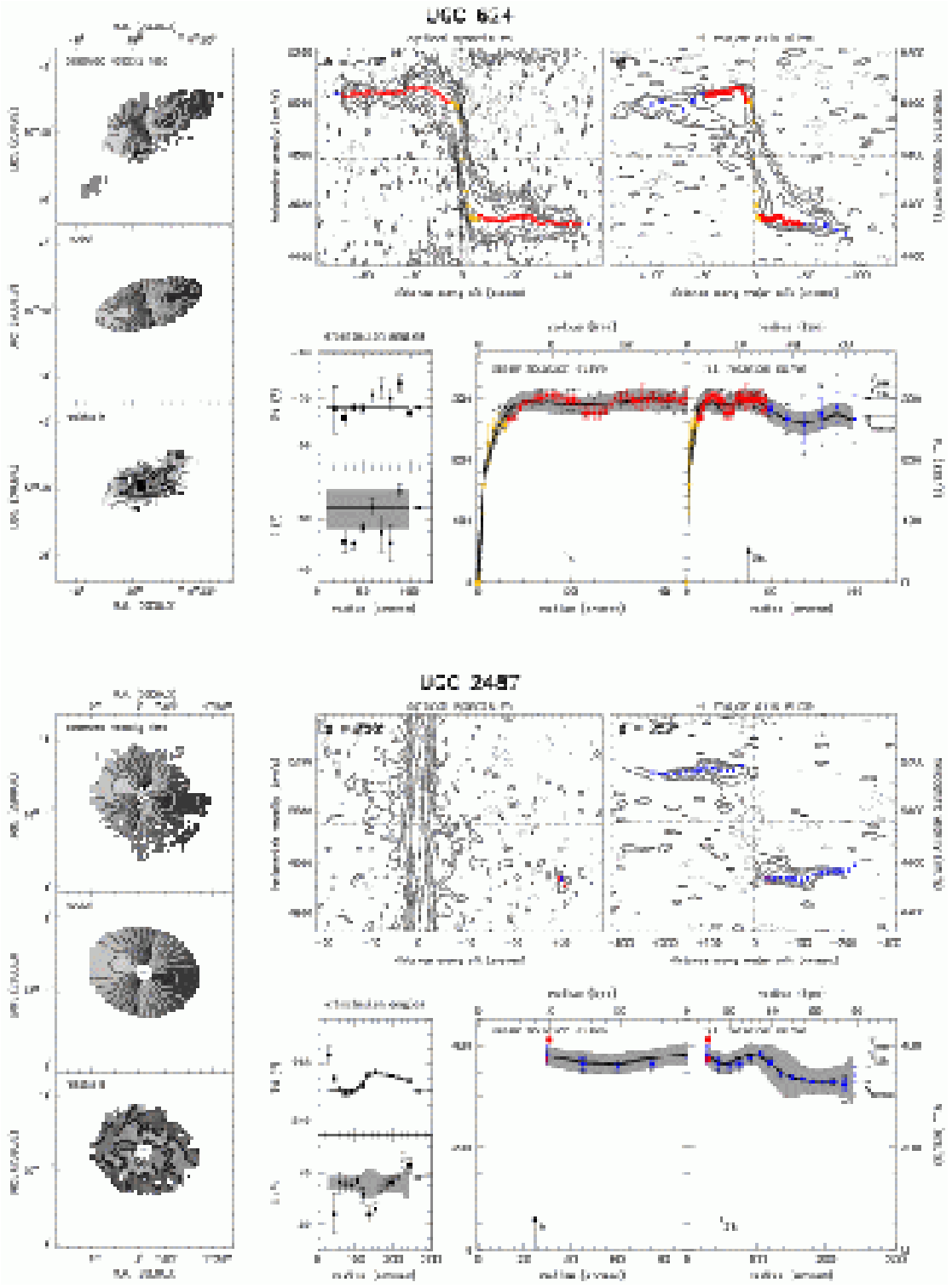,width=17.25cm}}
\end{figure*}

\end{document}